\documentclass{hcj-arxiv}
\usepackage{caption}
\usepackage{amsmath}

\usepackage{comment}

\usepackage{balance}  % to better equalize the last page
\usepackage{graphics} % for EPS, load graphicx instead 
\usepackage[T1]{fontenc}
\usepackage{txfonts}
\usepackage{mathptmx}
\usepackage{color}
\usepackage{booktabs}
\usepackage{textcomp}
\usepackage{microtype} % Improved Tracking and Kerning
\usepackage{ccicons}  % Cite your images correctly!
\usepackage{todonotes}

\usepackage{enumitem}
\usepackage{framed}

 \usepackage{booktabs}
 \usepackage{multirow}

\usepackage{balance}  % to better equalize the last page
\usepackage{graphics} % for EPS, load graphicx instead
\usepackage{times}    % comment if you want LaTeX's default font
\usepackage{url}      % llt: nicely formatted URLs

\usepackage{xspace}

\usepackage[T1]{fontenc}
\usepackage{textcomp}
\usepackage{graphicx} % for EPS use the graphics package instead
\usepackage{balance}  % for useful for balancing the last columns
\usepackage{booktabs} % for pretty table rules
\usepackage{ccicons}  % for Creative Commons citation icons
\usepackage{ragged2e} % for tighter hyphenation
\usepackage{dialogue}

\newenvironment{choruschat}
    {
        \begin{enumerate}[leftmargin=5pc,style=nextline,align=right]
        \small
        \sffamily
        \vspace{.3pc}
    }
    {
        \vspace{.3pc}
        \end{enumerate}
    }
\newcommand{\crowd}[1]{\item [\textbf{crowd}]\xspace \textbf{#1}} %crowd chat
\newcommand{\user}[1]{\item [\textbf{user}]\xspace #1} %user chat
\newcommand{\kenneth}[1]{{\small\textcolor{blue}{\bf [#1 --Ken]}}}

\newcommand{\ag}[2]
\usepackage{listings}
\usepackage{xcolor}

\colorlet{punct}{red!60!black}
\definecolor{background}{HTML}{EEEEEE}
\definecolor{delim}{RGB}{20,105,176}
\colorlet{numb}{magenta!60!black}

\lstdefinelanguage{json}{
    basicstyle=\normalfont\ttfamily,
    numbers=left,
    numberstyle=\scriptsize,
    stepnumber=1,
    numbersep=8pt,
    showstringspaces=false,
    breaklines=true,
    frame=lines,
    backgroundcolor=\color{background},
    literate=
     *{0}{{{\color{numb}0}}}{1}
      {1}{{{\color{numb}1}}}{1}
      {2}{{{\color{numb}2}}}{1}
      {3}{{{\color{numb}3}}}{1}
      {4}{{{\color{numb}4}}}{1}
      {5}{{{\color{numb}5}}}{1}
      {6}{{{\color{numb}6}}}{1}
      {7}{{{\color{numb}7}}}{1}
      {8}{{{\color{numb}8}}}{1}
      {9}{{{\color{numb}9}}}{1}
      {:}{{{\color{punct}{:}}}}{1}
      {,}{{{\color{punct}{,}}}}{1}
      {\{}{{{\color{delim}{\{}}}}{1}
      {\}}{{{\color{delim}{\}}}}}{1}
      {[}{{{\color{delim}{[}}}}{1}
      {]}{{{\color{delim}{]}}}}{1},
}

\newcommand{\system}{InstructableCrowd\xspace}

\journalyear{2019}
\journalvolume{6}
\journalissue{1}
\journalpages{113-146}
\articledoi{10.15346/hc.v6i1.7}
\journalcopyright{\copyright{} 2019, Huang, Azaria, Romero \& Bigham. CC-BY-3.0}

\title{InstructableCrowd: Creating IF-THEN Rules for Smartphones via Conversations with the Crowd}
\author{Ting-Hao K. Huang\affil{Pennsylvania State University}
        \and Amos Azaria\affil{Ariel University}
        \and Oscar J. Romero\affil{Carnegie Mellon University}
        \and Jeffrey P. Bigham\affil{Carnegie Mellon University}
        }
\authorrunning{T.-H.~K.~Huang, A.~Azaria, O.~J.~Romero and J.~P. Bigham}

\begin{document}

\maketitle

\begin{abstract}
Natural language interfaces have become a common part of modern digital life. Chatbots utilize text-based conversations to communicate with users; personal assistants on smartphones such as Google Assistant take direct speech commands from their users; and speech-controlled devices such as Amazon Echo use voice as their only input mode.
In this paper, we introduce {\em \system}, a crowd-powered system that allows users to program their devices via conversation.
The user verbally expresses a problem to the system, in which a group of crowd workers collectively respond and program relevant multi-part IF-THEN rules to help the user. 
The IF-THEN rules generated by \system connect relevant sensor combinations ({\em e.g.}, location, weather, device acceleration, {\em etc.}) to useful effectors ({\em e.g.}, text messages, device alarms, {\em etc.}).
Our study showed that non-programmers can use the conversational interface of \system to create IF-THEN rules that have similar quality compared with the rules created manually.
\system generally illustrates how users may converse with their devices, not only to trigger simple voice commands, but also to personalize their increasingly powerful and complicated devices.
%\kenneth{similarly, abstract only allows 150 *WORDS*, now is like 180+ something.}
\end{abstract}

\section{Introduction}
Intelligent personal computing devices -- such as smartphones, smartwatches, digital assistants ({\em e.g.,} Amazon's Echo) and wearables ({\em e.g.,} Google Glass) -- have become ubiquitous in society due to the power and convenience they offer. These devices are useful as shipped, but getting the most out of them requires tailoring them to their owner's preferences and needs. For example, after buying a smartphone, the user will usually first spend time customizing it by changing the wallpaper or adjusting the home screen layout. The same behavior is seen with nearly all other electronic devices, including personal assistants, tablets, laptops, and digital cameras. 
A great deal of customization takes place when the device is new, but the tuning process also usually continues at a slower pace over time as users adjust their devices in response to changing needs, the availability of new software or functionality, or shifts in personal circumstances. For example, a new security threat may lead to installing better firewall software; a near-miss with severe weather may prompt the user to change local weather alert preferences; and moving to a new city may lead to changing the parameters on travel or map software to reflect the user's new location. Users manually adjust the \textit{long-term behavior} of their devices in order to better fit their own behavior. 

As important as customization is, however, it is often held back by a variety of user difficulties. One is that devices are becoming ever more complicated: new features and capabilities provide power and flexibility, but at the cost of complexity. Customizing a device often requires the user wading through complex, multi-layer menus, searching for the right app, or experimenting with poorly explained settings. All of this can be confusing and intimidating. Furthermore, getting the most from a device usually requires programming it to react intelligently to events and automate responses, and many users find programming to be difficult and even frightening. The complexity of devices also means that even a small adjustment to a system's long-term behavior through programming could result in unintended consequences to the user's experience (when compared with simple, one-time interactions such as setting an alarm). Thus, the more complex the interaction with the device, the more important are high accuracy and robustness in understanding the user's needs.

\begin{figure}[t]
  \centering
  %\includegraphics[width=1.2\columnwidth]{img/workflow.jpg}
  %\hspace{-1pc}
  \includegraphics[width=0.9\columnwidth]{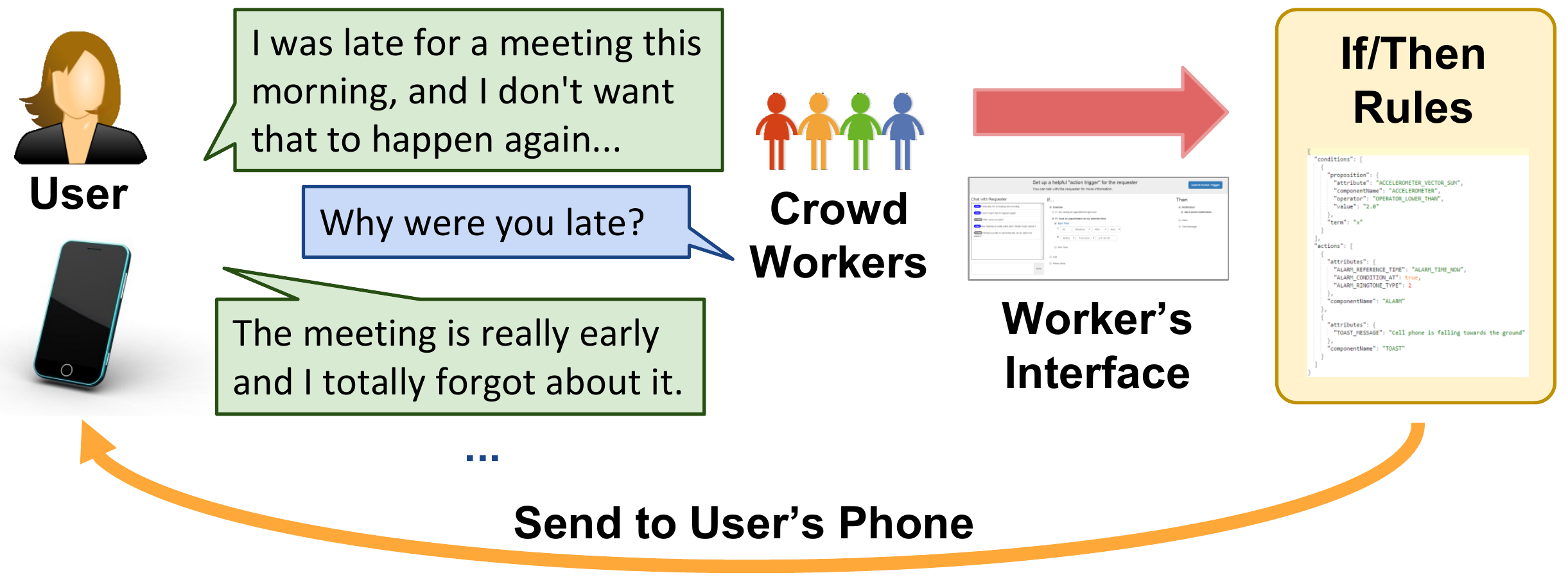}
  \caption{Users have a conversation with InstructableCrowd to create If/Then rules that then run on their phone to solve problems. The backend system is run by synchronous crowd workers who respond to the user, ask follow up questions, and create rules. Users can then review the rules on their phone to make sure they were what they wanted.}
  \label{fig:diagram}
  %\vspace{-.6pc}
\end{figure}

One technology has significant potential for addressing these problems is \textit{natural language interface.} Users could much more easily customize and even automate their devices if they could simply speak to them rather than wading through instruction manuals, menu trees, and tutorials. And in fact, natural language interfaces have become a common part of modern digital life already. Chatbots utilize text-based conversations to communicate with users; personal assistants on smartphones such as Google Assistant take direct speech commands from their users; and speech-controlled devices such as Amazon Echo use voice as their only input mode.

In this exploratory project, we introduce {\em \system}, a crowd-powered system that allows users to \textit{program} their devices and thus change their longer-term behavior via a natural language interface. \system is based around two key design decisions that address the main problems with device customization and automation outlined above. First, we have focused on creating relatively simple programs that are easy to use. Second, we make use of crowd workers to operate the natural language interface instead of using automated systems, since humans are much better at understanding and interpreting complex user requirements than current electronic systems.

Our programming system is oriented around relatively simple \textit{IF-THEN rules}, also known as \textit{trigger-action} rules. Modern smart devices, especially smartphones, contain a wealth of \textit{sensors} and \textit{effectors} that can be combined to perform useful customized tasks for their users. For example, they could be used to go beyond simple, static programming (such as setting a wake-up alarm to go off at a specific time every weekday) to customizations that are based on inputs and status information (like adjusting a wake-up alarm based on traffic conditions). 

A prominent example of this type of rule-based system is the mobile application IFTTT (If This Then That, ifttt.com). The service enables users to author simple trigger-action rules that contain only \textit{one} trigger ({\em e.g.}, a post on Twitter) and \textit{one} action ({\em e.g.}, synchronizing the latest Twitter post to Facebook)~\cite{ifttt_web}. The service is obviously useful --- it has millions of users~\cite{ifttt_web} --- and its simplicity makes it easy to use. However, that same simplicity also means that the system fails to cover many real-world scenarios~\cite{IftttMentalModel,daniel2012developing,SmartHomeProgramming2014}. Research has shown that 22\% of behaviors that people came up with require more than one sensor or effector~\cite{SmartHomeProgramming2014}. The complexity of rules people would like to create is likely to only increase as services like IFTTT continue to be integrated with other services and more devices. Therefore, in this project we focus on an extended version of IFTTT-style rules, in which the IF and THEN can each contain more than one sensor/effector. 

With the awareness of the limitation of automated dialog systems, we developed a \textit{crowd-powered} conversational agent. {\em \system} allows end users to create rich, multi-part IF-THEN rules via conversation with the crowd (Figure~\ref{fig:diagram}). A group of crowd workers is recruited on demand to talk with a user and create rules based on the conversation. With intelligent workers on a rich desktop interface supporting users, the interface can be simplified into a familiar speech or text chat client, allowing the system to be used on the go via mobile and wearable devices. Furthermore, users can discuss their problems with the crowd and get feedback to refine their requests. Users may know their problems, but not know what solutions would best resolve them. The crowd can help users identify possible solutions that the user didn't even know existed, and then create the rules needed to implement them. \system then lets users edit and improve the created rules. Controlled experiments showed that users are able to create complex rules using \system.

Through \system, we introduce a new method for enabling end users to program complex interactions with the wealth of sensors and effectors on their smartphones and other devices, which may have broader implications for the future of programming with speech.

\section{Related Work}
\system is related to prior work on {\em (i)} end-user programming, {\em (ii)} crowd-powered conversational agents, and (iii) automatic IF-THEN rules generation.

\subsection{End-User Programming}

%intro + domains
\system builds upon the long history of research and products in end-user programming~\cite{lieberman2006end}, which aims at enabling non-programmers to author or compose their own applications.
Early works in this field started from database~\cite{hanson1993overview} and email management~\cite{mackay1989experienced},
and later gradually became more common as more and more senors and effectors became available to general users~\cite{Bolchini:2007:DSC:1361348.1361353,bronsted2010service,Brush:2011:HAW:1978942.1979249,dahl2011end}.
For instance, CoScripter allowed end-users to program scripts by demonstration~\cite{coscripter,trailblazer}.
CoScripter used its corpus of scripts to allow easier creation of new actions from mobile devices~\cite{coco};
Sikuli is another famous end-user programming project~\cite{Yeh:2009:SUG:1622176.1622213}. 
Sikuli allows users to take a screenshot of a GUI element ({\em e.g.}, a toolbar button) and then directly use it as an element in a programming script to control the GUI's behavior ({\em e.g.}, clock the button.)
%CoScripter was an end user programming system for the Web, whereas \system provides similar functionality for the smartphone platform which contains a number of different sensors and effectors.

Trigger-action programming is one simple model of end user programming that the user form a new functionality by combining pre-defined triggers (sensors of ``IF'') with pre-defined actions (effectors of ``THEN'').
Many solutions were proposed to realize trigger-action programming, such as using existing notations of business processes modeling (BPM) to represent rules~\cite{brambilla2012combining}, 
adopting an effective workflow to create rules~\cite{SimpleFlow2013,kokciyan2012user},
or solutions for domain-specific applications~\cite{daniel2012developing}.
The IFTTT project has had great success by simplifying the composition among two applications and providing a user-friendly workflow and interface on mobile phones.
The concept of IFTTT has also been extended and adopted for use in various other domain, such as smart home applications~\cite{SmartHomeProgramming2014,HomeRules-2015}, cross-device interactions~\cite{action-trigger-authoring-2015},
the Internet of Things~\cite{tuomisto2014simple}.

%multiple sensors -> no for modern mobile devices 
IFTTT only allows rules to be composed of a single trigger and a single action.
Several frameworks were proposed to support multiple triggers (IFs) and actions (THENs).
Dey et al. created an interface that users can drag and drop multiple sensors and effectors on a sheet to create new rules~\cite{dey2006icap}.
Huang et al.~\cite{IftttMentalModel} and Ur et al.~\cite{SmartHomeProgramming2014} both extended IFTTT's interface to allow users select more than one triggers or actions.
However, most of these works focused on the challenges in designing interfaces or workflows for creating a rule and examined their solutions with participants using full-size monitors and keyboards, such as via Amazon Mechanical Turk.
Only few works focused on issues raised by mobile devices when creating complex rules.
H{\"a}kkil{\"a} {\em et al.} created a trigger-action programming system, Context Studio, on the Series 60 Nokia mobile phone back in 2005~\cite{hakkila2005interaction}.
While the mobile devices and sensors used in Context Studio were outdated, this project provided some early insights of challenges we face today.
On the other hand, competitors of IFTTT, such as Tasker, Llama, AutomateIt, On\{X\}, Atooma, and Microsoft's Flow all aimed to support multiple IFs and THENs in their product.
However, none of these have achieved the same success as IFTTT. 
%\oscar{Should we mention why? Tasker's and Llama's learning curves can be a little steep, but what about Atooma that has a more user-friendy interface?}
%\kenneth{don't think we want to elaborate on this}

%However, neither of these products have

%are as widely in use as IFTTT.\kenneth{this sentence is not good. Help needed.}

%limitation of IFTTT
Limitations of user programming were also studied.
Daniel et al.~\cite{daniel2012developing} pointed out that mashups platforms aimed at non-programmers are either powerful but too hard to use, or easy but too simple to be practical.
Huang et al.~\cite{IftttMentalModel} studied the mental model of IFTTT users and found that users do not always correctly understand how a sensor/effector works, which causes errors in user-created rules.
Recent work has been proposed which uses crowdsourcing to build software~\cite{latoza2016crowdsourcing}.

\subsection{Crowd-powered Conversational Agents}

Personal intelligent agents are now available on most smartphones, {\em i.e.}, Google Now on Android, Siri on iOS, Cortana on Windows phones. Google Now is known for spontaneously understanding and predicting user's life pattern ({\em e.g.}, flight schedules, or ``time to go home''), and automatically pushing notifications.
Such agents can understand a number of speech commands to help users more easily access functionality.
However, most of these virtual personal assistant are limited in their ability to understand their users.
Google Now only reacts to certain fixed set of events, and users have no manner to extend its capability based on their own needs;
Siri and Echo can perform speech queries, but are not able to understand complex verbal instructions to perform actions on the user's behalf. 
Although Echo allows to execute scripted actions via third-party services such as IFTTT\footnote{Users can apply IFTTT on Amazon's Alexa manually: https://ifttt.com/amazon\_alexa}, it requires users to manually program these behaviors in advance.
On the other hand, \system gives users the direct control to define intelligent behaviors their smartphones should perform, and uses the crowd to create these behaviors with conversational interaction.
%\kenneth{move to intro}

In response to this situation, crowd-powered intelligent agents were proposed.
Chorus is a crowd-powered assistant that can hold intelligent conversations~\cite{Chorus} and has been deployed to public~\cite{huang2016there}.
Users speak to it, and it responds back quickly.
Chorus is powered by a dynamic group of crowd workers (recruited on-demand) who propose responses and vote the best ones through.
An incentive mechanism encourages workers to contribute useful responses.
Potential downsides of crowdsourcing are cost and latency~\cite{chorus2}.
Guardian automates parts of Chorus by having the crowd transition existing Web APIs (Application Programming Interfaces) to dialog systems~\cite{Guardian}; and Evorus creates a framework that automates Chorus over time~\cite{Evorus}.

One limitation of these systems is that either Chorus or Guardian can only \textit{say} something to the user, but not \textit{do} something based on the conversation.
\system pushes the boundaries of crowd-powered conversational systems by allowing users to perform actions beyond information inquiry.
For instance, while users can discuss with Chorus to figure a good price of a flight ticket or verbally ask Guardian to query Travel APIs, users can not via a conversation configure a notification alert that monitors the dynamics of the ticket price with either systems.
%In a higher level, 
Enabling users to create a piece of computer-executable program via conversations opens up the opportunities of verbally ``instructing'' devices to customize their behaviors. The fact that today's voice-enabled devices such as Amazon's Echo allows users to set up IF-THEN rules (e.g., IFTTT) via mobile apps manually suggests the real users' needs of customizing their devices. 
\system explores performing these customization using conversational interface.
A similar effort that pushes the paradigm of personal assistant toward using conversations to set up or trigger applications can also be found in recent industrial products such as Google Assistant.

Alternatively, conversational assistants powered by trained human operators such as 
Magic\footnote{Magic: https://getmagic.com/}, Fancy Hands\footnote{Fancy Hands: https://www.fancyhands.com/} and Facebook M have also appeared in recent years.

%\kenneth{cite HCOMP 2016 paper}

\subsection{Automatic IF-THEN Rules Generation}
Automatically translating a natural-language utterance into the form that computers can execute is a well-known task in natural language processing, which is referred to as \textit{language understanding} or \textit{semantic parsing}.
For instance, Artzi {\em et al.} used a grounded CCG (Combinatory Categorial Grammar) semantic parsing approach to map instructions such as ``at the corner turn left to face the blue hall'' to actions that the agent (virtual robot) can execute~\cite{artzi2013weakly}; and
NaturalJava aimed to use a natural language interface for creating, modifying, and examining Java programs~\cite{price2000naturaljava}.

%ACL 2015
Particularly for IFTTT rules, Quirk {\em et al.} collected 114,408 IF-THEN rules and their natural-language descriptions from the IFTTT website, and demonstrated the possibility of producing IF-THEN rules based on corresponding descriptive text~\cite{quirk2015language}.
Several follow-up work that proposed different approaches such as attention-enhanced encoder-decoder model~\cite{dong2016language}, using latent attention~\cite{liu2016latent}, or syntactic neural model~\cite{yin17acl} further improved the accuracy of IFTTT rule generation.
Under the context of conversational assistance, Chaurasia {\em et al.} created an automated dialog system that generates IFTTT rules by having a conversation with users~\cite{chaurasia2017dialog}.
With a Free User-Initiative setting (``a more realistic setting''), Chaurasia's system achieved an accuracy of 81.45\% in generating IFTTT rules.
However, this performance is still not sufficient for practical use, and none of prior work attempted to produce multi-part rules that are more complex than that of IFTTT.

\section{Instructable-Crowd}

\system is implemented as an Android mobile application (Figure~\ref{fig:overview-diagram}) for supporting end-users to converse with crowd workers and describe problems they encounter, such as \textit{``I was late for a meeting this morning, and I don't want that to happen again.''}
The crowd workers can talk with the user and use an interface to select \textbf{sensors (IFs)} and \textbf{effectors (THENs)} to create an \textbf{If-Then rule} in response to the user's problem.
The rules are then sent back to the user's phone. % and applied instantly.
For instance, if the user mentions having trouble with early morning meetings, the crowd can create the rule ``send a notification the night before a meeting'' for the user.
Furthermore, \system is also able to merge multiple rules sent by different crowd workers to form a more reliable final rule.
We describe the system architecture and implementation details in this section.

\subsection{Rules, Sensors, and Effectors}

% Please add the following required packages to your document preamble:
% \usepackage{booktabs}
% \usepackage{multirow}
\begin{table}[htbp]
%\small
\centering
\begin{tabular}{@{}llll@{}}
\toprule
\textbf{Sensor} & \textbf{Trigger} & \textbf{Trigger Description} & \textbf{Attributes (Input Type)} \\ \midrule
\multirow{2}{*}{Bus} & Current location & \begin{tabular}[c]{@{}l@{}}The bus is currently at\\ a certain stop:\end{tabular} & \begin{tabular}[c]{@{}l@{}}Bus Number (Text)\\ Bus Stop (Text)\end{tabular} \\ \cmidrule(l){2-4} 
 & Future location & \begin{tabular}[c]{@{}l@{}}The bus will arrive at\\ a certain stop in minutes:\end{tabular} & \begin{tabular}[c]{@{}l@{}}Bus Number (Text)\\ Will Arrive at Stop (Text)\\ In How Many Minutes (Text)\end{tabular} \\ \midrule
\multirow{3}{*}{Calendar} & Current event & \begin{tabular}[c]{@{}l@{}}If I am having an event\\ right now that:\end{tabular} & Event Type (Select) \\ \cmidrule(l){2-4} 
 & \begin{tabular}[c]{@{}l@{}}Future event\\ (absolute time)\end{tabular} & \begin{tabular}[c]{@{}l@{}}If I will have an event that\\ (absolute time):\end{tabular} & \begin{tabular}[c]{@{}l@{}}Day (Select)\\ Start Time (Time)\\ End Time (Time)\\ Event Type (Select)\end{tabular} \\ \cmidrule(l){2-4} 
 & \begin{tabular}[c]{@{}l@{}}Future event\\ (relative time)\end{tabular} & \begin{tabular}[c]{@{}l@{}}If I will have an event that\\ (relative time):\end{tabular} & \begin{tabular}[c]{@{}l@{}}In How Many Minutes (Text)\\ Event Type (Select)\end{tabular} \\ \midrule
Call & Receive a call & If I receive a phone call that: & From (Text) \\ \midrule
Clock & Current time & The current time is: & \begin{tabular}[c]{@{}l@{}}At/Before/After (Select)\\ Time (Time)\end{tabular} \\ \midrule
Email & Receive an email & If I receive an email that: & Sent By (Text) \\ \midrule
\multirow{2}{*}{GPS} & Current location & I am currently located at: & Location Name (Text) \\ \cmidrule(l){2-4} 
 & \begin{tabular}[c]{@{}l@{}}Distance to\\ a location\end{tabular} & \begin{tabular}[c]{@{}l@{}}If my distance to a\\ certain location that:\end{tabular} & \begin{tabular}[c]{@{}l@{}}To (Text)\\ Is Greater/Less Than/Equals To (Select)\\ Distance (Text)\end{tabular} \\ \midrule
Message & Receive a message & If I receive a text message that: & \begin{tabular}[c]{@{}l@{}}Sent By (Text)\\ Contains the word(s) (Text)\end{tabular} \\ \midrule
News & Receive a news & If I receive a breaking news that: & Title contains the word(s) (Text) \\ \midrule
\multirow{2}{*}{\begin{tabular}[c]{@{}l@{}}Phone\\ Body\end{tabular}} & Phone falls & If my phone is falling. & N/A \\ \cmidrule(l){2-4} 
 & Drive & If I am driving. & N/A \\ \midrule
Weather & Weather forecast & If the weather forecast that: & \begin{tabular}[c]{@{}l@{}}Day (Select)\\ Forecast (Select)\end{tabular} \\ \bottomrule
\end{tabular}
\caption{Sensors (IFs) with their Triggers and Attributes as implemented in \system.}
\label{tab:if-list}
\end{table}

% Please add the following required packages to your document preamble:
% \usepackage{booktabs}
\begin{table}[htbp]
%\small
\centering
\begin{tabular}{@{}llll@{}}
\toprule
\textbf{Effector} & \textbf{Action} & \textbf{Action Description} & \textbf{Attributes (Input Type)} \\ \midrule
Alarm & Set an alarm & Set an Alarm that: & \begin{tabular}[c]{@{}l@{}}Day (Select)\\ Time (Time)\end{tabular} \\ \midrule
Calendar & Add an event & Add an Event on my Calendar that: & \begin{tabular}[c]{@{}l@{}}Day (Select)\\ Start Time (Time)\\ End Time (Time)\\ Event Type (Text)\\ Event Title (Text)\end{tabular} \\ \midrule
Call & Dial a call & Call: & \begin{tabular}[c]{@{}l@{}}To (Text)\\ What to Say (Text)\end{tabular} \\ \midrule
Email & Send an email & Send Email(s) that: & \begin{tabular}[c]{@{}l@{}}To (Text)\\ Email Title (Text)\\ Email Content (Text)\end{tabular} \\ \midrule
Message & Send a message & Send Message(s) that: & \begin{tabular}[c]{@{}l@{}}To (Text)\\ Message Content (Text)\end{tabular} \\ \midrule
Notification & Send a notification & Push me a Notification that: & Notification Content (Text) \\ \bottomrule
\end{tabular}
\caption{Effectors (THENs) with their Actions and Attributes implemented in \system.}
\label{tab:then-list}
\end{table}

\begin{figure*}[htbp]
  \centering
  \includegraphics[width=0.8\textwidth]{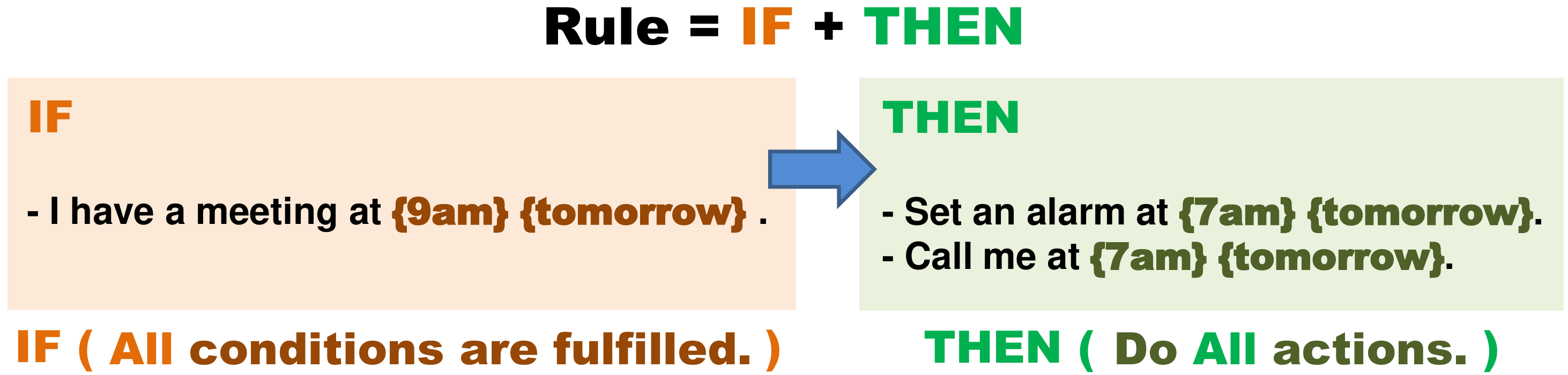}
  \caption{Example of a rule in \system. A Rule is defined as a tuple that contains an IF part and a THEN part. The IF part contains a set of Sensors that describe aspects of the user's life and context, and the THEN part contains a set of Effectors that can be performed.}
  \label{fig:rule}
\end{figure*}

%\kenneth{newly written section. proof reading needed.}
In this work, a \textbf{Rule} is defined as a tuple that contains an \textbf{IF} part and a \textbf{THEN} part.
The IF part contains a set of \textbf{Sensors} (also referred to as \textbf{IFs}) that describe aspects of the user's life and context.
For instance, the ``Calendar'' application describes the status of all calendar events of the user, and the ``Phone Body'' sensor describes the physical motions of the smart phone ({\em e.g.}, phone is moving).
Both can be Sensors in the IF part.
The THEN part contains a set of \textbf{Effectors} (also referred to as \textbf{THENs}) that can be performed, such as push a notification, set an alarm, and send a text message, etc.
It is noteworthy that InstructableCrowd allows more than one Sensors/Effectors in each part, while IFTTT only allows one.
An overview of an example rule is shown in Figure~\ref{fig:rule}.

Each Sensor has one or more Triggers that can be selected.
%, and each Effector also has one or more Actions can be performed.
For instance, the ``calendar'' sensor could have three different Triggers that reflect the status of 1) currently ongoing events, 2) future events at an absolute time ({\em e.g.}, 9am today), or 3) future events at a relative time ({\em e.g.}, in 30 minutes.)
Similarly, one Effector can also have one or more Actions to perform.
Each Trigger and Action is composed of a set of \textbf{Attributes} to specify the details of the condition.
For instance, for configuring ``Calendar'' sensor
%'s ``Future Event (Relative Time)'' Trigger,
to tell if the user has any events in 30 minutes with the ``Future Event (Relative Time)'' Trigger, the ``In How Many Minutes'' attribute needs to be filled with ``30,'' and the ``Event Type'' attribute needs to be filled with ``Any.''
In this paper, we focused on observing end-user and workers behavior in \textbf{selecting Sensors/Effectors} and \textbf{filling Attributes}.
%We considered Trigger/Action selection as an essential part of filling attribute values, instead of isolating it as a separated step. 

%the process of \textbf{Sensor/Effector Selection} and \textbf{Attribute Filling} and thus considered Trigger/Action selection as an essential part of filling attributte values.

The full list of Sensor and Effectors with their Triggers/Actions and Attributes used in our study are listed in Table~\ref{tab:if-list} and Table~\ref{tab:then-list}.

\subsection{Conversational Agent for the End-user}

\begin{figure*}[htbp]
  \centering
  \includegraphics[width=1\textwidth]{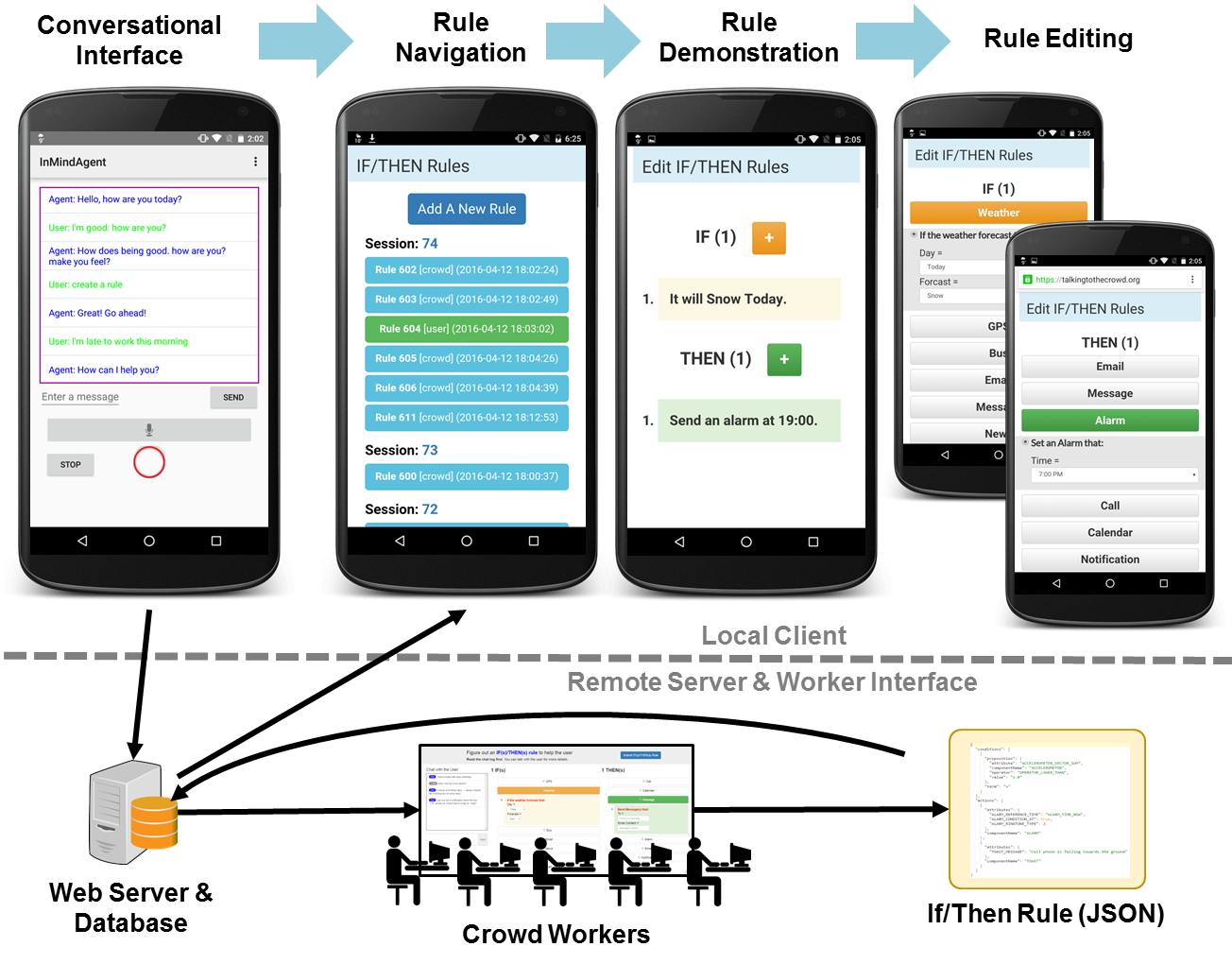}
  \caption{InstructableCrowd users have a conversation with crowd workers about a problem they are having. Crowd workers collectively create IF-THEN rules that may help the end user solve their problem using sensors and effectors available on the smartphone platform. The rules are then sent back to the user's phone for review, editing, and approval. The rules then run on the smartphone.}
  \label{fig:overview-diagram}
  %\vspace{-.5pc}
\end{figure*}

\system is implemented as a conversational agent for Android smartphones.
By calling the personal agent's name or clicking on the red button (as shown in Figure~\ref{fig:overview-diagram}), the user is able to give the agent commands via voice or text.
The client side records the user's speech and sends it to the server, which in turn sends this speech on to Google Automatic Speech Recognition; 
the user can also use text entry to input the command.
\system adopts the LIA framework~\cite{azaria2016instructable}, which uses a combinatory categorial grammar (CCG) parser to parse the input text into a logical form and execute the corresponding commands, 
to recognize user's voice input.
%perform language understanding and dialog management.
%The user is able to have simple conversations such as ``How are you?'' %or ``What's your name?'' with the agent, and 
Once the user give verbal commands such as \textit{``create a rule,''} LIA connects to \system and initiates the rule creation process.
%\kenneth{add text entry part, it's not clear}
%Once the agent receives the command, we use the framework described in~\cite{azaria2016instructable}, named LIA (Learning by Instruction Agent), to perform dialog management 
%execute the command.
%\kenneth{explain why LIA is here: why we need LIA if it's crowdsourcing?} \amos{I removed it for now, maybe bring it back if accepted}
%LIA uses a combinatory categorial grammar (CCG) parser to parse the input text into a logical form and execute the corresponding commands.
%\system recognizes verbal commands such as \textit{``create a rule''} to initiate the rule creation process. %InstructableCrowd also supports the execution of many different commands via the LIA framework \cite{azaria2016instructable}.

At the beginning of each conversation, \system posts 10 Human Intelligence Tasks (HITs) to Amazon Mechanical Turk to recruit a group of crowd workers.
Each worker will be directed to a web-based interface (Figure~\ref{fig:ui}), where they can view the user's messages, respond to the user, and compose an IF-THEN rule based on the user's request.
The user and workers communicate with each other synchronously via a web server (Figure~\ref{fig:overview-diagram}).
A similar system framework has been used by several real-time crowd-powered conversational agents, such as Chorus~\cite{huang2016there,Chorus} and Evorus~\cite{Evorus}.
%to talk with a user and create rules based on the conversation via a web-based worker interface (Figure~\ref{fig:ui}).
%With intelligent workers on a rich desktop interface supporting users, the interface can be simplified into a familiar speech or text chat client, allowing the system to be used on the go via mobile and wearable devices.

The user may then describe his problems and converse with the crowd to figure out which rules to create (the workers converse by text, and the user, may either use text or voice).
Once the rule is created, it is sent back to the user's phone, where a Decision Rule Engine component~\cite{romero:2017,romero:2018} will store, validate and process that rule.
Currently, the system is implemented and tested on the Android OS 6.0.1 and the server is implemented in Java.
%\kenneth{don't feel like OS v6.0.1 and Java is important details...}
%\jpb{Probably not, adding the detail probably doesn't hurt though, as long as that version is not too old}

\subsection{Rule Editor for the End-user}

\system also provides an editing interface for the user to manually create new rules, edit them and edit rules received from crowd workers.
As shown in Figure~\ref{fig:overview-diagram}, the user is able to navigate all received rules and click on each rule for additional details.
All rules are grouped together by the conversational session in which the rule was created.
Crowd-generated rules are blue, and the rules created or edited by the user are green.
%To enable users from different backgrounds to easily understand the conditions and actions of a 
In order to ease on the comprehension of these rules, we created a template-based natural language description for each Trigger.
For instance, the description template of ``Weather'' sensor's forecast Trigger is ``It will \texttt{\big[weather\big]} \texttt{\big[day\big]}.''
If ``Weather'' sensor's this trigger is selected, along with the ``Day'' attribute filled with ``Tomorrow'' and the ``Forecast'' attribute filled with ``Snow'', the displayed description will be ``It will Snow Tomorrow.''
%\footnote{The complete list of sensors and effectors with their attributes used in this paper are shown in Table~\ref{tab:if-list} and Table~\ref{tab:then-list}.}
On the editing interface, the description will be generated automatically in real-time and enable the user to quickly check the rule they just created or edited.
The user can also use this rule editor to manually create an IF-THEN rule from scratch on their phone without talking to the crowd.
In our user study, participants use various approaches to create IF-THEN rules with \system. % and we evaluated the results.
Our end-user editing interface is inspired by the IFTTT mobile APP. % IF.
However, it enables the user to combine multiple IFs and THENs while IFTTT focuses on one-to-one APP compositions. 
%As a result, it is difficult or impossible to create rules necessary to solve some problems, {\em i.e.}, those involving two or more IF conditions. Such rules are often necessary, {\em e.g.}, if I have a meeting soon and am more than 5 miles away, if it is time to leave work and my normal bus is late and my alternative bus is not late, etc.
%\kenneth{what is impossible mean}

%\kenneth{Add merged rules into the workflow}

%We manually created a natural language template for each sensor, and the attributes will be fill in automatically. 
%this display is real-time on the client side, and the users are able to check their editing results instantly. 
%This editor is an stand-alone section of the code, so users can also create a rule from scratch solely on their phone without talking to the crowd. 

%Our interface is inspired by IFTTT interface, however, enable an multiple APP selection and thus make it more tricky for both user and the crowd.

%In the future, the natural language description could also be generated by crowd worekers, but for now, we do it ourselves.
%Another minor issue here is the "naming" problem of the rule, when an rule get more complex (multiple APPs invloved), it's lack of a clear and simple way to automatically name each rule. In the future, we might enable users to name all rules.

\subsection{Worker Interface}

\begin{figure*}[t]
  \centering
  \includegraphics[width=.95\textwidth]{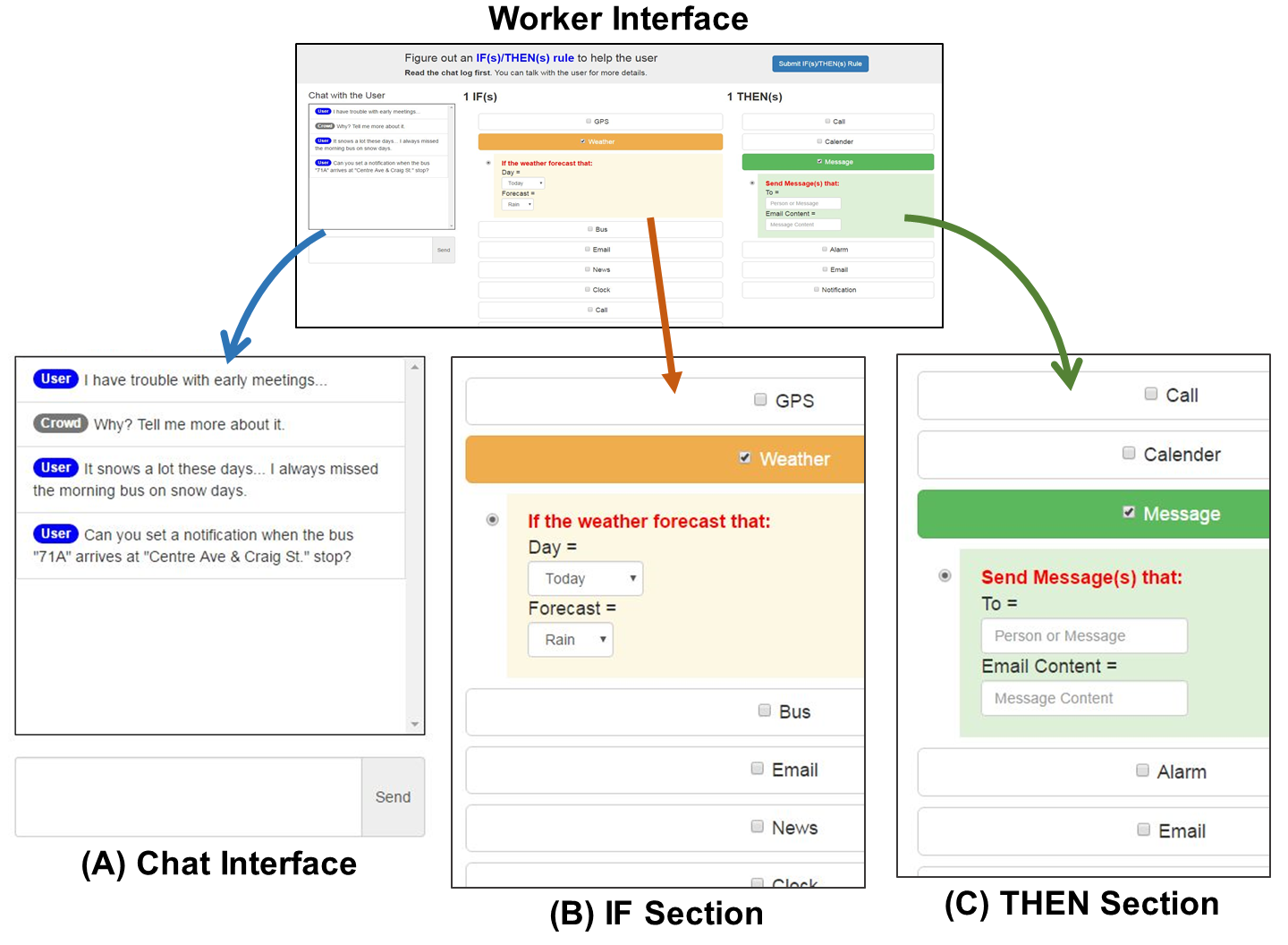}
  \caption{Worker interface. A chat interface (A) allows workers to talk to the end user to discuss the problem. The IF section (B) allows the worker to specify Sensors, along with Triggers (in red text) and their Attributes; the THEN section (C) allows them to specify Effectors, along with Actions (in red text) and their Attributes.}
  \label{fig:ui}
  \vspace{-1pc}
\end{figure*}

The worker interface allows crowd workers to select Sensors (IFs) and Effectors (THENs) easily.
The interface contains three main parts (Figure~\ref{fig:ui}).
1) The web-based chat interface allows workers to discuss the problem with the end-user in real-time. 
2) The IF section contains a set of sensors on the user's phone that describe aspects of the user's life and context.
%For instance, the ``Calendar'' describes the status of all calendar events of the user, and the ``Phone Body'' describes the physical motions of the smart phone ({\em e.g.}, phone is moving).
%Both are considered ``sensors'' in \system.
Workers first select appropriate Sensors ({\em e.g.}, Calendar) in the IF conditions, and then select Triggers under the Sensors ({\em e.g.}, Future Event (Relative Time)), and finally fill in appropriate attribute values ({\em e.g.}, In How Many Minutes = 30.)
3) The THEN section allows workers to select Effectors and corresponding Actions, and fill in attribute values.
%Effectors are the actions that can be performed on user's smart phone such as push a notification, set an alarm, and send a text message, etc.
By selecting IFs and THENs, the worker is able to create rules that trigger certain actions based on specific conditions.

\subsection{Merge Multiple Crowd-Created Rules by Voting}
\label{sec:crowd-vote}
%\kenneth{Newly written. Polish needed.}
\system recruits multiple workers for each conversation; therefore, multiple rules are received respectively from each conversation.
End users are free to pick any rules submitted by the crowd, or
wait until the rules merged automatically into a \textit{final rule}.
%wait the rules are merged together automatically into a \textit{final rule}.
Our automated rule-merging process uses output agreement to identify the best components to use.
First, any Sensors and Effectors that are selected by more than 2 workers (our current threshold) are included in the final rule.
%\oscar{maybe we can explain why we use this method and based on which criteria. It sounds to me similar to Borda Count Voting used as single-winner election method, so we can include citations.}
%\kenneth{added citation}
%\amos{Why 2? Doesn't this depend on the number of workers? Maybe just say some threshold, and leave the details for the results section?}
Second, for each Sensor/Effector picked in the first step, its Trigger/Action that is selected by most workers will be chosen.
Finally, for each selected Trigger/Action,
\system fills each attribute with the value that was proposed by the most workers.
If two values were proposed by an identical number of workers, \system selects the value which was proposed earliest.
Output-agreement mechanisms such as ESP Game for collecting image labels~\cite{von2004labeling} have been widely used to obtain reliable human-generated labels from multiple workers~\cite{GameWithAPurpose2008}.
Its variation, input-agreement, has also been introduced~\cite{Law:2009:INM:1518701.1518881}.
%The voting process is inspired by 
%\kenneth{unfinished}

\subsection{Modular Sensors (IF) \& Effectors (THEN)}

We designed a general JSON (JavaScript Object Notation) schema to represent each sensor and effector.
The rules created by the crowd are represented as a combination of sensors and effectors in this JSON format.
New sensors and effectors can thus be added easily.
For example, the following is the Weather sensor's JSON file representing that ``it will snow tomorrow'' (Trigger = Weather forecast).

\vspace{.5pc}

\begin{lstlisting}[language=json,firstnumber=1]
{
  "name": "if-weather",
  "condition": "if-weather-forecast",
  "attributes": [
    {
      "name": "if-weather-forecast-day",
      "value": "Tomorrow",
      "type": "select"
    },
    {
      "name": "if-weather-forecast-condition",
      "value": "Snow",
      "type": "select"
    }
  ]
}
\end{lstlisting}

%"condition": {
%    "proposition": {
%        "componentName": "CALENDAR",
%        "attribute": "CALENDAR_START_TIME",
%        "operator": "OPERATOR_TIME_EQUAL",
%        "value": "9:30",
%        "referenceAttribute": "CALENDAR_TOMORROW"
%    }
%}

\vspace{.5pc}

The following is the JSON representation of the Alarm effector for ``set the alarm at 7am tomorrow''  (Action = Set an alarm.) 

\vspace{.5pc}

\begin{lstlisting}[language=json,firstnumber=1]
{
  "name": "then-alarm",
  "condition": "then-alarm-send",
  "attributes": [
    {
      "name": "then-alarm-send-day",
      "value": "tomorrow",
      "type": "text"
    },
    {
      "name": "then-alarm-send-time",
      "value": "07:00",
      "type": "text"
    }
  ]
}
\end{lstlisting}

%"action": {
%    "componentName": "ALARM",
%    "attributes": {
%        "ACTION_TYPE": "ALARM",
%        "ALARM_REFERENCE_TIME": "ALARM_TIME_NOW",
%        "ALARM_RINGTONE_TYPE": 2
%      }
%}

\vspace{.5pc}

The following is the JSON representation for an IF-THEN rule ``IF it will snow and I have a meeting at 9am tomorrow, THEN set alarm at 7am,'' which includes 2 sensors (Weather and Calendar) and 1 effector (Alarm.)
%includes a list of sensor items and effector items. 
%\jpb{you might include a concrete example of an interesting rule here, instead of a generic psuedocode one}
%\oscar{I modified the order, so ruleId goes first, then conditions and finally actions}
%It snowed last night. I was late for work this morning and missed an important meeting at 9am because I had to take care of all the snow. My boss was quite upset and warned me this can not happen again. (Intermediate scenario.)

\vspace{.5pc}

\begin{lstlisting}[language=json,firstnumber=1]
{
  "if": [
    {
      "name": "if-weather",
      "condition": "if-weather-forecast",
      "attributes": [
        {
          "name": "if-weather-forecast-day",
          "value": "Tomorrow",
          "type": "select"
        },
        {
          "name": "if-weather-forecast-condition",
          "value": "Snow",
          "type": "select"
        }
      ]
    },
    {
      "name": "if-calendar",
      "condition": "if-calendar-future",
      "attributes": [
        {
          "name": "if-calendar-future-day",
          "value": "Tomorrow",
          "type": "select"
        },
        {
          "name": "if-calendar-future-type",
          "value": "Meeting",
          "type": "select"
        },
        {
          "name": "if-calendar-future-start",
          "value": "09:00",
          "type": "time"
        }
      ]
    }
  ],
  "then": [
    {
      "name": "then-alarm",
      "condition": "then-alarm-send",
      "attributes": [
        {
          "name": "then-alarm-send-day",
          "value": "tomorrow",
          "type": "text"
        },
        {
          "name": "then-alarm-send-time",
          "value": "07:00",
          "type": "text"
        }
      ]
    }
  ]
}
\end{lstlisting}

%{
%    "ruleID": "rule_1",
%    "conditions": [
%        {condition_1}, ..., {condition_n}
%    ],
%    "actions": [
%        {action_1}, ... , {action_n}
%    ]
%}

\vspace{.5pc}

%These rules and actions are made possible by a general architecture that we have built to allow systems to access a list of sensors and effectors, and then specify what sensor conditions should lead to what actions.

New sensors and effectors can be added easily once they are implemented in our middleware, by simply adding new JSON entries for them.
Currently, we implemented 10 sensors and 6 effectors in \system (Table~\ref{tab:if-list} and Table~\ref{tab:then-list}.)
As we go forward, we plan to continue expand the set of available sensors/effectors.

%\kenneth{<- do we need to say we're going to expand?...} \jpb{Yes, i think that's fine. i would repeat how many of each there are now, because i think it's actually pretty impressive}

%\kenneth{How's the back-end working? How hard it can be to create an back-end rule hub on Android? What do we learning? -- Rule validation?}

\subsection{Decision Rule Engine}

%We developed a middleware framework that allows the communication and integration between the front-end (the user interface) and the back-end (the server-side processes and the sensors which retrieve information from third party web-services, e.g. weather).
%The middleware's purpose is manifold: 1) to provide access to a set of well-defined services (e.g., calendar, weather, news, search, audio/video streaming, activity recognition, etc.); 2) to provide access to a set of sensors and effectors (e.g., location, accelerometer, battery, gyroscope, send sms, send emails, etc.); 3) to mediate communication amongst UI components and services through a Message Broker component which validates, transforms, routes and aggregates all kind of messages that are sent over those components; 4) to track and monitor all the user interaction with apps and system components; 5) to validate crowd rules and trigger their corresponding actions (effectors); and 6) to give support to high-level decision-making by the agent.

\begin{figure}[htbp]
  \centering
  \includegraphics[width=.8\columnwidth]{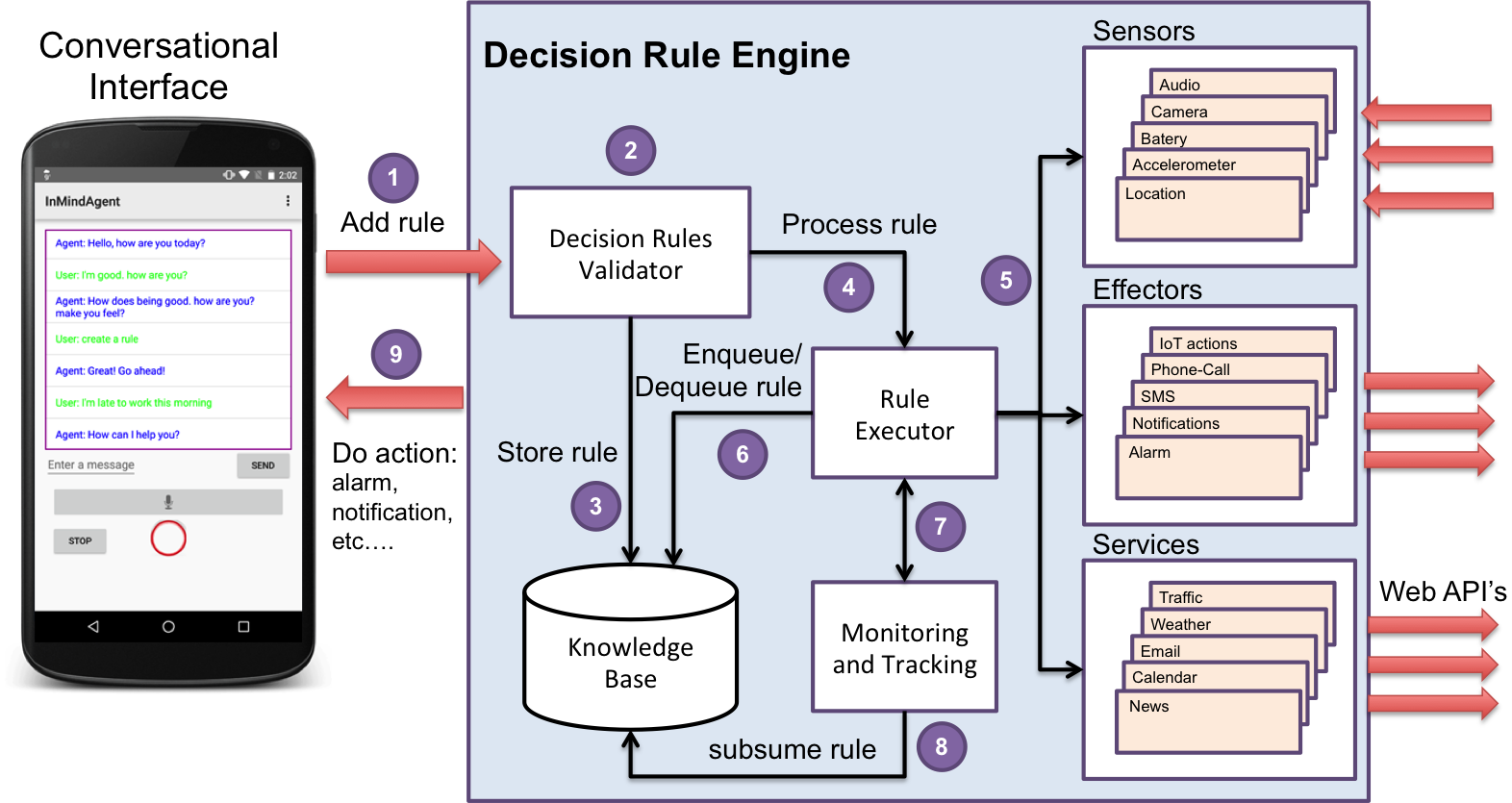}
  \caption{The architecture of \system's Decision Rule Engine. The step (1) to step (9) outlines the work flow, message passing and how Decision Rule Engine components cooperate over time to manage rules created by user and crowd-workers.}
  \label{fig:rule-engine}
\end{figure}

%\kenneth{i took out the abbreviation ``DRE''. readers don't need to know this...}
The Decision Rule Engine is in charge of validating, storing, processing and executing rules created by either a crowd-worker or the user.
Decision Rule Engine is composed of multiple modules that interact with each other in order to execute an action given a set of specific conditions that are true.
These modules are interconnected as shown in Figure~\ref{fig:rule-engine}.
The following outlines the work flow (in steps), message passing and how Decision Rule Engine components cooperate over time to manage rules created by user or crowd-workers.

%\kenneth{Written by Oscar and slightly modified by Kenneth. Need proofreading.}

\begin{itemize}

    \item \textbf{[Decision Rule Validator]} After the user or crowd worker has defined a new rule to be added (Step 1 in Figure~\ref{fig:rule-engine}), this component validates the syntax of that rule according to the sensors' and the effectors' attributes and constraints (Step 2).
    For instance, if the rule has a condition that refers to attribute \texttt{<CALENDAR\_START\_TIME>}, the validator will parse this condition and check that in fact there exists a sensor called ``Calendar'' that has an attribute called \emph{startTime}, which must be of type \emph{Date} and whose value must be a date/time that occurs later than current date/time.
\vspace{.5pc}    
    \item \textbf{[Knowledge Base]} Once the rule is parsed and validated, it is stored in a knowledge base where can be accessed anytime by any component (Step 3). These rules are stored locally for performance and privacy reasons, so potentially sensitive information contained within the rule is protected.
 \vspace{.5pc}   
    \item \textbf{[Rule Executor]} After validation, the rule is immediately processed in order to determine whether it should be executed in that moment (Step 4). If so, it invokes actions from the appropriate effectors (Step 5). If not, it adds the rule to a queue so it can be executed later when all its conditions are true. The Rule Executor periodically checks to see if each enqueued rule needs to be executed (Step 6).
\vspace{.5pc}
    \item \textbf{[Monitoring \& Tracking]} This module is responsible for monitoring the rule execution process (Step 7) by checking if there are rules that are either never triggered or conflicting with each other ({\em e.g.}, one rule intends to turn the GPS on while the other one intents to turn it off.)
    When conflicts occur, the Monitoring/Tracking module temporarily subsumes the less relevant rule ({\em i.e.}, the one that has been activated less frequently) and then user is asked to confirm this subsumption decision (Step 8).
    \vspace{.5pc}
    \item \textbf{[Built-in \& External Sensors/Effectors]} In addition to built-in sensors and effectors that are part of the operating system, such as GPS and SMS Messages, some virtual sensors/effectors are based on external services, such as the Weather forecast and News feeds. In our implementation, we use a RESTful API to upload, extract and collect information from web servers.
    %Sensors and effectors are phone's built-in input and output controls.
    %Sometimes, a control may play both roles as sensor and effector. For instance, the SMS control behaves as sensor when it receives an incoming message and as an effector when it sends out a message. Sensors always match rule triggers whereas Effectors always match rule actions.
    %We are also using \emph{Services} to extend far beyond the set of sensors and effectors available on a smart phone.
    %These services use REST APIs to upload, extract and collect information from web servers such as Weather forecast, News feeds, etc. 
    %When and how these Sensors and Effectors interact with each others is determined through the decision rules we have described above.
    Finally, user is always aware of action execution through notifications, text messages, alarms, etc. (Step 9).
    
\end{itemize}

%The Rule Validator module plays a fundamental role during the creation and validation of crowd rules. 
%The Rule Validator receives a rule creation request in the JSON format (this rule was either created by the crowd or by the user).
%In general, its goal is to validate the conditions over all the IFs, and if the validation is true, it triggers the relevant THENs in the rule.
%Whereas \textit{Sensors} always match rule conditions and \textit{Effectors} always match rule actions.

\section{User Study}
For evaluating the performance of \system, we conducted a set of in-lab user study.
Our goal is to understand if creating IF-THEN rules using conversation would sacrifice rule quality, compared with using a graphic user interface (GUI).
%For this purpose, 
%\kenneth{re-write this opening?}
%For evaluating the capability of \system in terms of the accuracy and time required to create IF-THEN rules
%We conducted a study to understand whether conversing with \system would effectively allow users to create rules on %par with what they could do with a GUI interface on their phone.
%We believed that \system would not be necessarily better than the GUI interface, but that if it performed nearly as well could be useful in contexts in which using a GUI was difficult or undesired.
%This is similar to how speech-driven interfaces are often used for tasks for which there are equivalent GUI applications.
%Nevertheless, people often prefer natural language.
Furthermore, we specifically recruited non-programmers because one of the benefits of using \system is that complex rules can be created without the need for a programming-like interface. Participants created rules using a mobile application in a control condition to allow us to compare with how users currently create rules using applications such as IFTTT.

\subsection{Scenario Design}

%To design a set of scenarios that will be used in the lab-based study, we first conducted a formative study on Amazon Mechanical Turk (MTurk).
%We posted the following three questions and collected answers from one hundred crowd workers:
%\kenneth{Use the IFTTT mturk paper to motivate our Scenario design} 

We designed the following 6 scenarios (S1 to S6) inspired by~\cite{IftttMentalModel}, along with a gold-standard set of sensors and effectors for each that we consider to be ground truth for assessing the performance.\footnote{
The attributes which were not specified in a gold-standard rule indicate that the user or worker should leave these attributes blank.
In the evaluation, the textual attributes such as message content or email content will be examined manually.
It is also noteworthy that in this section we only listed one common gold-standard rule, while more than one rule ({\em e.g.,} adding or alternating notifications) could be considered valid for a scenario.
%More gold-standard rules ({\em e.g.,} adding or alternating notifications) could be included when evaluating the system.
We describe the details of evaluation in Section~\ref{sec:tech-eval}.
} We further categorized scenarios into three difficulty levels based on the numbers of sensors and effectors the scenario requires.
S1 and S2 are \textbf{easy} scenarios (1 sensor and 1 effector), 
S3, S4, and S5 are \textbf{intermediate} scenarios (2 sensors and 1 effector), and
S6 is \textbf{hard} scenario (2 sensors and 2 effectors).

%\jpb{missing from these are the slots that need to be filled and their values, e.g., for S1 the News needs to contain the keywored ``Steelers''}
%\kenneth{I don't think we should list ALL the gs-values, especially when the effectors can have various different values. Given we've listed all the sensors and effectors in Table~\ref{tab:if-list} and Table~\ref{tab:then-list}, I think it should be good?}\jpb{I would list them - like, IF: News (keyword: Steelers), or whatever}

\begin{enumerate}

\item \textbf{[S1] Sports:} I am very interested in the performance of the ``Steelers'' and would like to get an immediate notification if there is a news article mentioning them. (Easy scenario.)
%\textit{(IF: News, THEN: Notification. Easy scenario.)}
\vspace{.3pc}
\begin{itemize}
    \item \textit{\textbf{IF}: News (Receive a news: Title contains the word(s) = ``Steelers'')}
    \item \textit{\textbf{THEN}: Notification (Send a notification: Notification Content = ``News of Steelers!'')}
\end{itemize}
\vspace{.5pc}
\item \textbf{[S2] Message:} My mother likes to send me text messages. I work in a restaurant so I cannot reply to her messages very often at work. However, my grandfather was hospitalized last week and my mother is taking care of him now. I do not want to miss any important message about my grandpa. (Easy scenario.)
%\textit{(IF: Message, THEN: Notification. .)} 
\vspace{.3pc}
\begin{itemize}
    \item \textit{\textbf{IF}: Message (Receive a message: Sent By = Mom, Contains the word(s) = ``grandfather'')}
    \item \textit{\textbf{THEN}: Notification (Send a notification: Notification Content = ``Mom just texted you a message about grandfather!'')}
\end{itemize}
\vspace{.5pc}
\item \textbf{[S3] Snow \& Meeting:} It snowed last night. I was late for work this morning and missed an important meeting at 9am because I had to take care of all the snow. My boss was quite upset and warned me this can not happen again. (Intermediate scenario.)
\vspace{.3pc}
\begin{itemize}
    \item \textit{\textbf{IF}: Weather (Weather forecast: Day = today, Forecast = snow) + Calendar (Future event [absolute time]: Day = tomorrow, Event Type = meeting, Start Time = 09:00)}
    \item \textit{\textbf{THEN}: Alarm (Set an alarm: Day = tomorrow, Time = 07:00)}
\end{itemize}
%\textit{IF: Weather (Weather forecast: Day = ``today'', Forecast = ``snow'') + Calendar}\\
%\textit{THEN: }
\vspace{.5pc}
\item \textbf{[S4] Drive \& Call:} I just heard that a large percentage of car accidents are caused by talking on the phone while driving. I decided I am not going to answer any phone calls while driving. Therefore, when I am driving, if anyone calls me, I would like to automatically reply to him/her with a message saying ``Sorry I'm driving.'' (Intermediate scenario.) %\textit{(IF: Phone Body for Driving + Message, THEN: Message. Intermediate scenario.)} 
\vspace{.3pc}
\begin{itemize}
    \item \textit{\textbf{IF}: Phone Body (Drive) + Call (Receive a call: From = Anyone)}
    \item \textit{\textbf{THEN}: Message (Send a message: To = People mentioned in ``IF(s)'', Message Content = ``Sorry, I am driving.'')}
\end{itemize}
\vspace{.5pc}
\item \textbf{[S5] Bus:} I usually leave work after 5pm and take Bus ``53'' home at the ``Washington St.'' stop. However, the ``53'' buses are not common. I prefer not to wait at the bus stop unless the bus is coming soon. It takes me about 5 minutes to walk from my office to the ``Washington St.'' stop, and it also takes about 5 minutes for Bus ``53'' to drive from the ``Hamilton St.'' stop to the ``Washington St.'' stop. (Intermediate scenario.)
%\textit{(IF: Bus + Clock, THEN: Notification. Intermediate scenario.)} 
\vspace{.3pc}
\begin{itemize}
    \item \textit{\textbf{IF}: Bus (Current location: Bus Number = 53, Bus Stop = ``Washington St'') + Clock (Current time: At/After/Before = After, Time = 17:00)}
    \item \textit{\textbf{THEN}: Notification (Send a notification: Notification Content = ``Bus 53 will be arriving at Washington St. stop soon!'')}
\end{itemize}
\vspace{.5pc}
\item \textbf{[S6] Late for Dinner:} My wife Amy does not like me to be late home when we have a big scheduled dinner. So, if I am going to have a big dinner at home in 30 minutes, but I am still far away -- say, 30 miles -- from home, please send Amy a message saying ``I might be home late''. Also, give a phone call to ``Ben's Flower Shop'' and tell them to ``Prepare a small surprise bouquet.'' (Hard scenario.)
%\textit{(IF: GPS + Calendar, THEN: Message + Call. Hard scenario.)} 
\vspace{.3pc}
\begin{itemize}
    \item \textit{\textbf{IF}: GPS (Distance to a location: Is Greater/Less Than/Equals To = Is Greater Than, To = Home, Distance = 30) + Calendar (Future event [relative time]: Event Type = Dinning, In How Many Minutes = 30)}
    \item \textit{\textbf{THEN}: Message (Send a message: To = Amy, Message Content = ``I might be home late.'') + Call (Dial a call: To = Ben Flower Shop, What to Say = ``Prepare a small surprise bouquet for me.'')}
\end{itemize}
\end{enumerate}

In our post-study survey, we asked participants to rate how realistic these scenarios are, in the scale of 1 (very unrealistic) to 7 (very realistic).
The mean rating among the twelve participants was 6.25 (SD=0.62).

%In our user study, 

\begin{figure}[t]
  \centering
  \includegraphics[width=0.9\columnwidth]{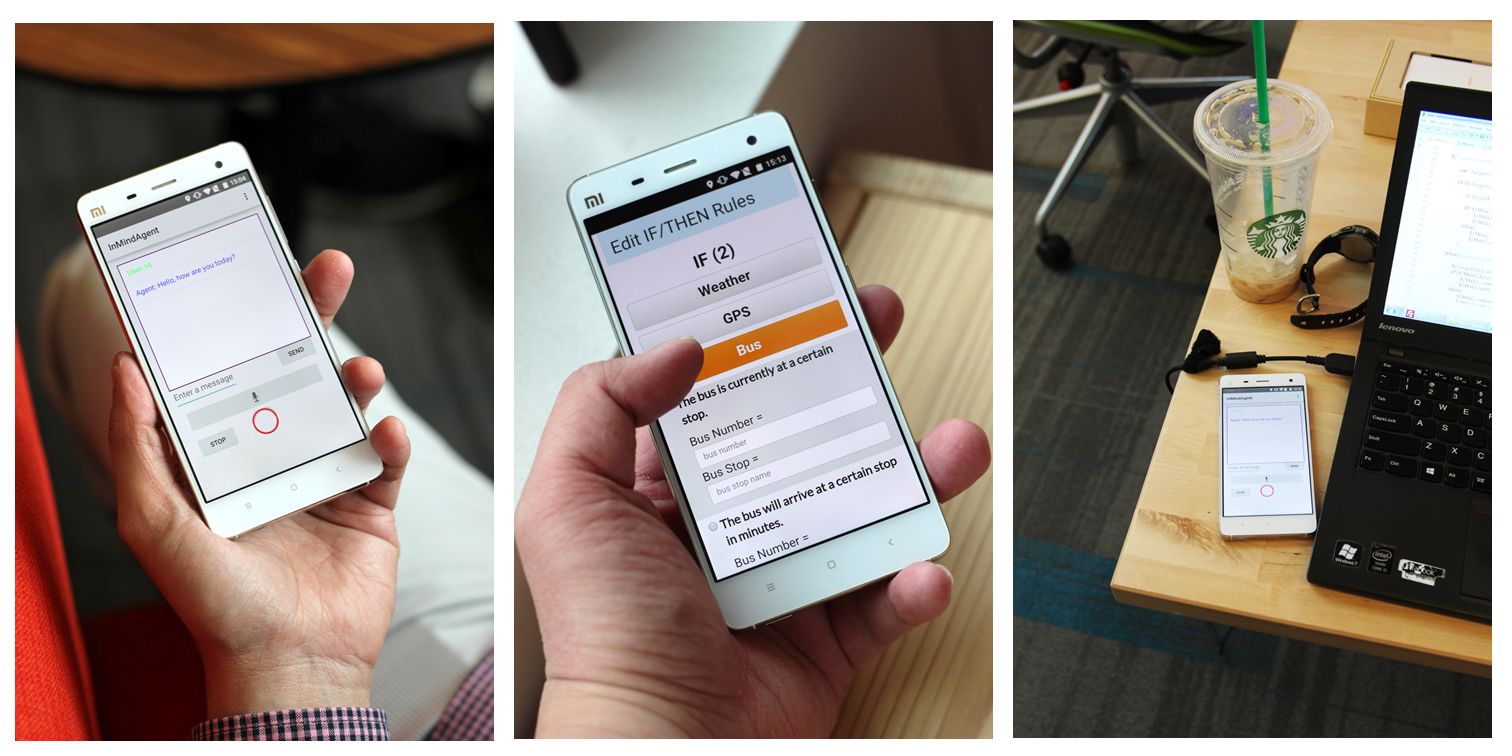}
  \caption{User study setting. While waiting for responses from the crowd, %(less than 5 minutes), 
  participants used their own laptops or mobile devices to simulate the likely context of use in the real world.}
  %\vspace{-1pc}
  \label{fig:user-study}
\end{figure}

\subsection{User Study Setup}

We conducted a lab-based user study in which we asked participants to create an IF-THEN rule for each scenario using one of the following conditions:

\begin{enumerate}
    \item \textbf{[Condition 1] \system}: The participant first talks to the crowd via \system (using text or voice, depends on the participant's preference) and waits to receive rules submitted from the crowd workers. %Once upon starting to receive rules, 
    The participant then selects a rule that they prefer and manually edits it to create the final rule. 
    Each conversation was shown to 10 workers, and each worker creates an IF-THEN rule based on the conversation, respectively.
    \vspace{.5pc}
    \item \textbf{[Condition 2] User}: The participant uses the rule editor on the phone (as shown in Figure~\ref{fig:user-study}) to manually create a rule.
\end{enumerate}

In condition (1), three data points were recorded: 
the crowd-created rule that was picked by the participant (which we refer to as \textbf{Crowd Only}),
the rule edited by the participant (\textbf{Crowd + User}),
and the rule that was created by merging all ten crowd-created rules (\textbf{Crowd Voting}) using the process described in Section~\ref{sec:crowd-vote} (threshold for including a sensor/effector was 2.)
We refer to condition (2) as \textbf{User Only}.

%Each participant was asked to create rules for all 6 scenarios, 

%Note that in condition (1) we can obtain two rules in total, one directly from the crowd (which we refer to as ``Crowd Only''), and the other from user's editing (which we refer to as ``Crowd + User'').
%If the user thinks the crowd-created rule is effective and does not perform editing, we still consider the crowd-created rule as the product of ``Crowd + User'' setting.

%In the case that the user thinks the crowd-generated rule is effective and thus does not edit

For recruiting participants, we posted the information on social media sites such as Facebook and Twitter.
We also posted flyers on the campus of Carnegie Mellon University (Pittsburgh campus) and University of Pittsburgh. 
The goal of this project is to enable users to compose applications for their own usage, especially for the users who do not know how to program.
Therefore, we recruited participants which had very limited experience in programming or none at all.
%In the posted information, we explicitly invited people who have none or very limited experience in programming.
People who volunteered to participate our study were directed a Web form for signing up, in which we asked people to self-report their programming skill level (``How good are you at programming?''), from 1 (``I don't know anything about programming.'') to 7 (``I'm an expert programmer.'').
%At the end of the recruiting period, 
We selected the earliest 14 participants who signed up with a self-reported programming skill level of 1 or 2.
The first 2 participants were recruited for the pilot study, in which we tested and refined our study protocol and the system, and the remaining 12 participants were recruited for the formal user study.
All the results reported in this paper were based on the formal user study with these 12 participants, who were aged from 26 to 36 years (mean = 29.42, SD = 3.48); 8 are female and 4 are male; and 11 participants rated their own programming skill level as 1 (out of 7), and only one participant self-rated as 2 (out of 7).
It is noteworthy that the goal of this project is to examine the feasibility of using a natural language interface to create IF-THEN rules. While our participants were of a younger population, we believe that a user study with twelve participants is sufficient to show the idea of \system works, and that \system can be helpful to some users.

%12 participants (25-36 years old) were recruited locally.

In our user study, we scheduled a one-hour time slot with each participant and brought them in the lab, respectively. 
Each participant was requested to create an IF-THEN rule which would resolve each of the 6 scenarios.
The participants were asked to solve three scenarios via \system (condition 1), and three other scenarios via the rule editor (condition 2).
The scenarios were controlled for the condition they were associated with. That is, each scenario was given to 6 subjects as condition 1 and to 6 other subjects as condition 2. In addition, the scenarios were controlled for the order in which they appeared, that is, each scenario was given in each possible order (first, second, third, fourth, fifth and last) exactly once for each condition. This was done in order to reduce the learning-effect.
Participants were instructed to follow the scenarios as close as possible, but were allowed to propose minor changes during the conversation, {\em e.g.}, change ``send me notification'' to ``send me an email.''
%Participants are free to choose text entry or voice entry with speech recognition to converse with the crowd.
Participants were also free to use their own laptop or mobile devices when they waited for the response from the crowd (as shown in Figure~\ref{fig:user-study},) because we believe this setting is more realistic for users who try to converse via instant messaging on mobile devices.
A post-study questionnaire was used to collect subjective feedback from the participants.
The compensation for each participant was \$20.

For each conversational session, \system posted a HIT (Human Intelligence Task) with 10 assignments to MTurk.
The price of each assignment was \$0.50 USD. 
%\amos{So is this a total of \$5 per participant?} \kenneth{no. each trial costs \$5}
During a conversational session, multiple workers could communicate with the participant via their interface and submit rules respectively.
156 unique workers on MTurk participated in our experiments. 
%\amos{why not 12*10=120?} \kenneth{workers can do multiple trials if they want.}
%Worker's interface was implemented as an on-line chat room that multiple workers can talk to the user at the same time.
%If multiple rules were received from multiple crowd workers in the same session,
%participants are instructed to select their favorite rule to edit.
All sessions, chats, and rules were recorded in a database with timestamps.
We also timed how long the participant took to create each rule by using the rule editor.

%\kenneth{one more level: component setting?}
As listed in Table~\ref{tab:if-list} and Table~\ref{tab:then-list},
in the user study crowd workers and end-users had 10 sensors to choose from:
Email, Bus, Message, GPS,
Weather, Call, Clock, Calender, News, and Phone Body (for driving and phone falling);
and 6 effectors:
Message,
Email,
Alarm,
Call,
Notification, and
Calendar (for adding an event).
%The details of each sensors and effectors are listed in Table~\ref{tab:if-list} and Table~\ref{tab:then-list}.
%Each sensor had an average of 1.6 attributes to fill in (SD=1.1), and each effector had an average of 2.5 attributes (SD=1.4).
%Note that not all of these sensors/effectors were implemented and working with the middleware.
% -> Don't feel like we need to say this haha
%\kenneth{say sensor conditions here!}

\section{Rule Quality Evaluation}
\label{sec:tech-eval}
%\kenneth{two opening paragraphes of this section is new. Need proofreading.}

%We also performed a technical evaluation of \system in order to \jpb{I don't really understand what the technical evaluation is...}.

In this section we evaluated the quality of resulting rules in each setting.
%, and the \textit{time} that users spent to create these rules.
%\kenneth{ok technical eval is mislearning. So what should we call it? }
In order to assess the quality of a composed IF-THEN rule,
%In technical evaluation 
we focused on two subtasks: \textbf{sensor/effector selection} and \textbf{attribute filling}.
Composing an IF-THEN rule contains three sub-tasks:
sensor/effector selection, 
trigger/action selection,
and attribute filling.
For instance, to effectively know that you have an early meeting tomorrow, the ``Calendar'' sensor firstly needs to be selected, and then its ``Future Event (Absolute Time)'' trigger needs to be selected, and finally the ``Start Time'' attribute needs to be filled with ``Before 8am.''
Since each sensor used in our study on average only has 1.5 triggers (SD=0.71) and each effector only has 1 action, we did not evaluate the performance of trigger/action selection separately, but merge it as a part of attribute filling.
Namely, in the case that the triggers/actions selected by users or the crowd were incorrect, we noted the accuracy of attribute filling as zero in this sensor/effector.

In this section, we describe the evaluation results of \system and demonstrate that the system is able to produce high-quality IF-THEN rules via conversation. %with a minor cost of longer user active time.

\subsection{Evaluation of Sensor/Effector Selection}

The evaluation process was as follows:
First, we expanded the set of our original gold-standard rules to include participant-created rules which were useful, but not exactly what we anticipated.
For instance, in S3, some participants decided to send emails to the boss at work instead of setting up an earlier alarm; in S2, one participant decided to reply to his/her mom with a message instead of setting a push notification.
We went through all the submitted rules and added the effective solutions that we did not think of initially.
Second, we allowed extra or alternative effectors if appropriate.
For instance, some participants thought that setting a push a notification is not enough and decided to send an email or to set an alarm.
We considered these alternative rules are also effective.
Finally, a piece of software was created to perform an automated evaluation on all recorded rules.

Selecting a set of correct sensors/effectors from a pool of candidate is a \textit{retrieval} task.
We therefore use precision, recall, and F1-score to evaluation this sub-task. These values are calculated as follows. % (The effectors in THEN are evaluated in the same manner.)
%We compared the generated rule's sensors and effectors against the gold-standard sensors and effectors of the given scenario.
%We calculate the precision and the recall of the sensors in the IF as follows:
\begin{equation*}
\text{Precision} = \frac{|\{\text{Selected Sensors}\}\cap\{\text{Gold-Standard Sensors}\}|}{|\{\text{Selected Sensors}\}|}
\end{equation*}
\begin{equation*}
\text{Recall} = \frac{|\{\text{Selected Sensors}\}\cap\{\text{Gold-Standard Sensors}\}|}{|\{\text{Gold-Standard Sensors}\}|}
\end{equation*}
\begin{equation*}
\text{F1-score} = \frac{2 \times \text{Precision} \times \text{Recall}}{\text{Precision}+\text{Recall}}
\end{equation*}

%The F1-score is the harmonic mean of precision and recall. 
When a rule is partially correct, we selected the gold-standard rule which results in the highest F1-score to report the numbers in this paper.
%When multiple gold-standard rules exist, the gold-standard rule which results in the highest F1-score was used to report the numbers in this paper.
%\amos{What does this mean? If there are several possible answers, aren't they all supposed to be considered correct?}
%\kenneth{Yes. But when a rule is only partially correct, I picked the gs-answer give the highest F1-socre to calculate the performance.(text updated)}
The overall evaluation results are shown in Table~\ref{tab:seneor-select}.
Both ``Crowd+User'' and ``Crowd Voting'' settings achieved comparable performances to that of the ``Crowd Only'' setting is both IF and THEN parts.
Selecting correct sensors in IF is harder than selecting correct effectors in THEN, which is expected due to the tolerant nature of our evaluation setup for THEN.
We observe that ``Crowd Voting'' resulted in a higher average recall, which suggested that a group of crowd workers is, collectively, less likely to forget picking some sensors than an individual user.
We also notice that participants actually corrected errors in the crowd-created rules, as both the average precisions and recalls are higher in ``Crowd+User'' than ``Crowd Only''.
%of both the IF and THEN were improved after user editing.
For instance, in the ``Late for Dinner'' scenario (S6), one common mistake was that crowd selected only one of Calender or GPS sensors, instead of both.
Two different participants fixed this error by adding back the missing sensor.
Another similar example occurred in the ``Bus'' scenario (S5), where the crowd sometimes missed the ``Clock'' sensor which can indicate the current time is after 5pm.
One participant fixed this by adding the Clock sensor back to the IF.

% Please add the following required packages to your document preamble:
% \usepackage{booktabs}
\begin{table}[htbp]
\centering
\begin{tabular}{@{}lrrrrrrr@{}}
\toprule
 & \multicolumn{3}{c}{\textbf{IF}} & \multicolumn{3}{c}{\textbf{THEN}} & \multicolumn{1}{c}{\textbf{Avg}} \\ \midrule
 & Precision & Recall & F1 score & Precision & Recall & F1 score & F1 score \\
\textbf{User Only} & 0.94 & 0.85 & 0.89 & 0.98 & 0.99 & 0.98 & 0.94 \\
\textbf{Crowd Only} & 0.94 & 0.77 & 0.85 & 0.97 & 0.90 & 0.94 & 0.89 \\
\textbf{Crowd+User} & 0.94 & 0.83 & 0.89 & 1.00 & 0.94 & 0.97 & 0.93 \\
\textbf{Crowd Voting} & 0.92 & 0.89 & 0.91 & 0.95 & 0.96 & 0.96 & 0.93 \\ \bottomrule
\end{tabular}
\caption{Sensor/Effector selection overall performance. Both ``Crowd+User'' and ``Crowd Voting'' settings achieved comparable performances to that of the ``Crowd Only'' setting is both IF and THEN parts.}
\label{tab:seneor-select}
\end{table}

\begin{figure}[t]
  \centering
  %\vspace{-.2pc}
  \includegraphics[width=0.9\columnwidth]{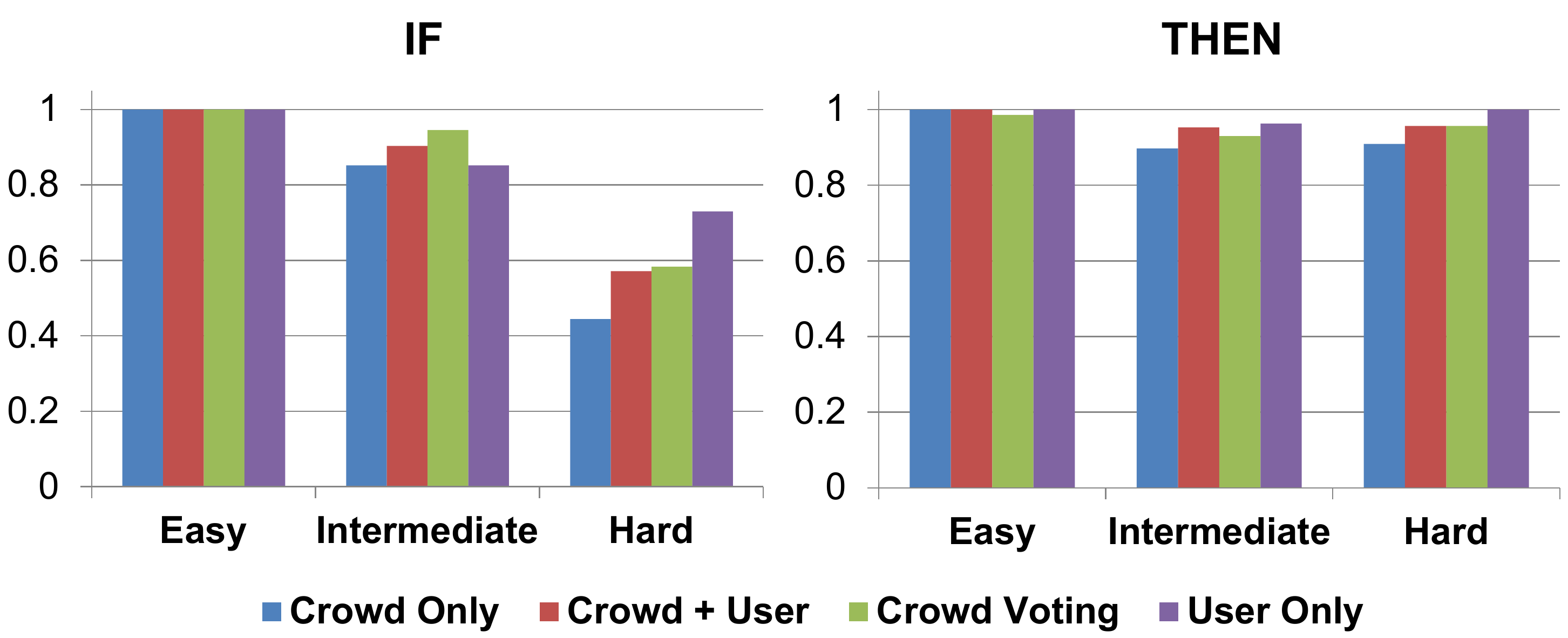}
  \caption{Average F1-score of sensor/effector selection in easy, intermediate, hard scenarios. ``Crowd Voting'' performed similarly or slightly better than ``User Only'' in easy and intermediate rules, but worse in hard rules.}
  %\vspace{-.5pc}
  \label{fig:sensor_selection_3_level}
\end{figure}

We also evaluated the performance based on the scenarios' difficulty level.
The dynamics of F1-scores are shown in Figure~\ref{fig:sensor_selection_3_level}.
While the THEN parts were not influenced much, the F1-scores in IF parts' decreased as the scenarios got more complex.
``Crowd Voting'' performed similarly or slightly better than ``User Only'' in easy and intermediate rules, but worse in hard rules.
These results also indicate the number of sensors and effectors influences the difficulty level of composing the rule, while other factors such as abstraction level and type of sensors/effectors also reportedly play important roles~\cite{SmartHomeProgramming2014}.

%\kenneth{What to say here? Why hard is hard even it only has 1 IF}
%First, when conditions are complex. It's harder even the \#sensor is the same.
%Second, in the hard case, user is still very good.

%\kenneth{Error analysis here?}

\subsection{Evaluation of Attribute Filling}

% Please add the following required packages to your document preamble:
% \usepackage{booktabs}
\begin{table}[htbp]
%\small
\centering
\begin{tabular}{@{}lrrr@{}}
\toprule
 & \textbf{IF} & \textbf{THEN} & Avg \\ \midrule
\textbf{User Only} & 98.3\% & 95.0\% & \textbf{96.7\%} \\
\textbf{Crowd Only} & 81.4\% & 90.0\% & 85.7\% \\
\textbf{Crowd + User} & 89.2\% & 93.3\% & \textbf{91.3\%} \\
\textbf{Crowd Voting} & 86.4\% & 95.0\% & \textbf{90.7\%} \\ \bottomrule
\end{tabular}
\caption{Attribute filling overall performance. While the ``Crowd Voting'' setting achieved the same average accuracy as that of the ``User Only'' in the THEN part, its average accuracy is lower than ``User Only'' in the IF part.}
\label{tab:att-filling}
%\vspace{-1pc}
\end{table}

\begin{figure*}[t]
  \centering
  \includegraphics[width=\textwidth]{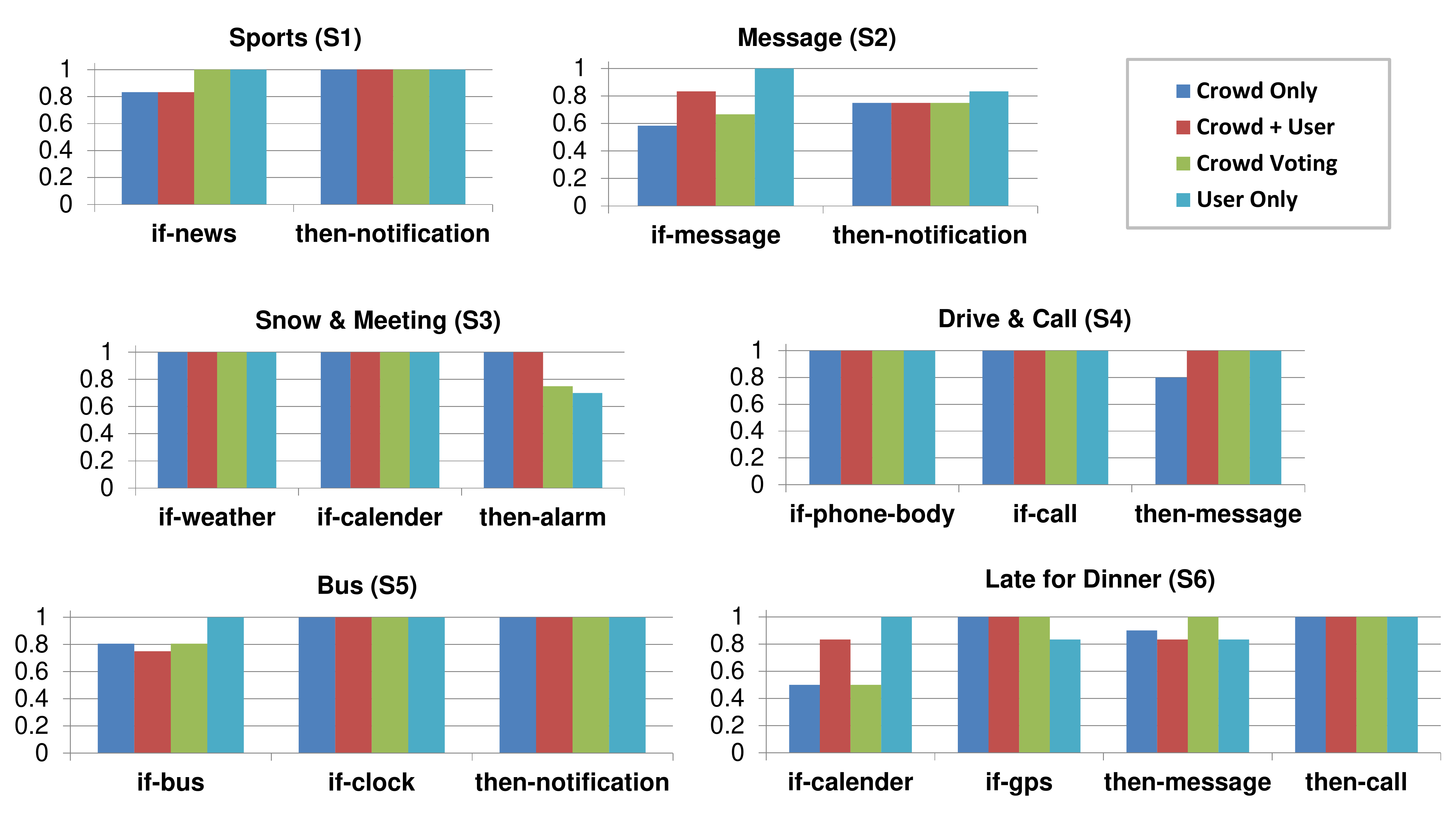}
  \caption{Average accuracy of attribute filling of correctly-selected sensors/effectors. ``Crowd Voting'' performed similarly as ``User Only'' in most cases. We analyzed S2, S5, and S6 and found that crowd errors are mainly caused by communication gap and misunderstanding of attributes.
  %\jpb{also try to find some takeaways to emphasize here about these results, also make sure these charts are readable in black and white printing}\kenneth{can't think of a good take away here...}
  }
  %\vspace{-.5pc}
  \label{fig:att-filling-detail}
\end{figure*}

The evaluation process of attribute filling is similar to that of sensor/effector selection.
Any value for an attribute which seemed appropriate was considered to be correct.
For instance, the content of the sent messages or emails could vary, and we manually labeled the effectiveness of each ``content'' attribute in effectors;
the ``Day'' attribute (Table~\ref{tab:if-list}) in the Weather sensor of S3 could be set to either ``Today'' or ``Tomorrow'', however, it would only be judged as correct if the Alarm's ``Day'' attribute (Table~\ref{tab:then-list}) was set to the same value.
Software was created to evaluate these attributes automatically.

For a given sensor/effector $S$ that is correctly selected, we calculate the accuracy of its attribute values as follows.
\begin{equation*}
\text{Accuracy} = \frac{\text{Number of Attributes in $S$ with correct values}}{\text{Number of Attributes in $S$}}
\end{equation*}
\begin{equation*}
\text{If trigger/action of $S$ is incorrect, Accuracy}=0.
\end{equation*}

The overall evaluation results of attribute filling are shown in Table~\ref{tab:att-filling}.
While the ``Crowd Voting'' setting achieved the same average accuracy as that of the ``User Only'' in the THEN part, its average accuracy is lower than ``User Only'' in the IF part.
To understand the sources of this performance gap, we analyzed the average accuracy of attributes in each sensor/effector of each scenario, as shown in Figure~\ref{fig:att-filling-detail}.
We observed the sensors (IF) where ``Crowd Voting'' resulted in a lower accuracy than that of ``User Only'' (i.e., the Message sensor in S2, the Bus sensor in S5, and the Calendar sensor in S6) and identified two sources of crowd workers' errors:
\textbf{communication gap} and \textbf{misunderstanding the meanings of triggers}.
%mixed up with other attributes
One source of the errors was the communication gap between the end-user and crowd workers.
Namely, the user falsely expressed or missed some information when talking to the crowd.
For instance, in S2, one participant falsely said ``dad'' often sent him/her messages (instead of ``mom''), and the crowd therefore filled ``dad'' in the ``Sent By'' attribute;
in S5, %user 9
one participant did not mention to the crowd that it usually takes 5 minutes to walk to the bus stop, so the crowd arbitrarily filled the ``In How Many Minutes'' attribute of Bus sensor with 2 minutes (trigger = ``Future location'').
Another source of the errors is the misunderstanding the meanings of triggers.
In S6, we found that some crowd workers confused the ``Future Event (absolute time)'' trigger with ``Future Event (relative time)'' trigger of the Calendar sensor.
%, and arbitrarily made up an absolute time ({\em e.g.}, ``5pm'') instead of assigning a relative time (e.g., ``in 30 minutes.'')
In addition, both users and crowd workers have typos in their attributes.
For instance, a worker misspelled ``Steelers'' as ``Stelers'' in S1, and another worker entered ``19:00'' as the ``start time of the meeting'' in S3, while the expected answer is ``07:00.''

\section{User Active Time}
\begin{figure}[htbp]
  \centering
  %\vspace{-.2pc}
  \includegraphics[width=0.95\columnwidth]{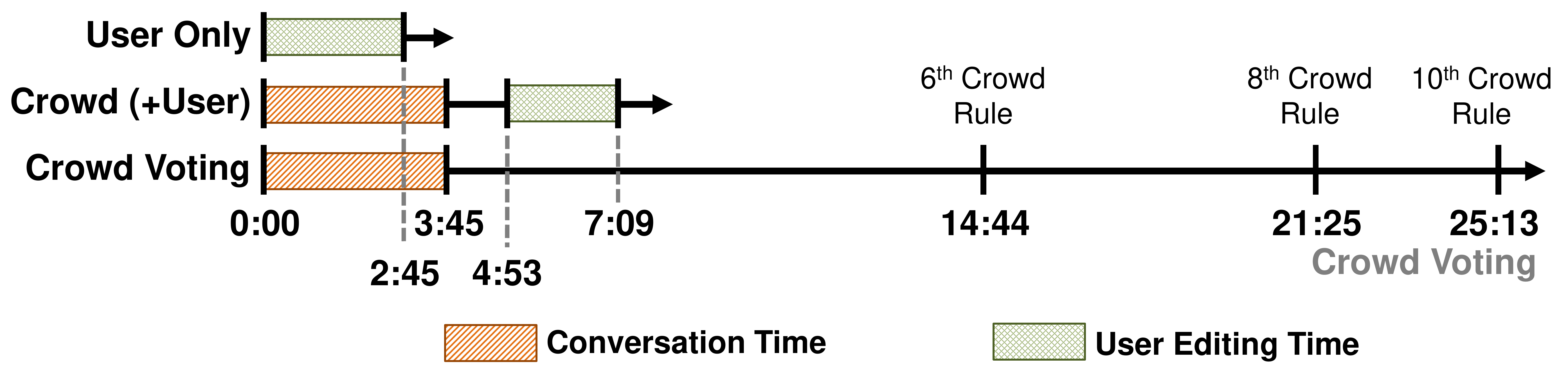}
  \caption{The complete timeline of \system. With the cost of a slightly longer user active time, \system is able to generate rules with comparable quality user-created rules. Furthermore, in our post-study survey (Section~\ref{sec:user-feedback}) the participants who preferred using \system over rule editor claimed that \system is ``faster'' or ``quick'', while their user active time of using \system is actually longer.}
  %\vspace{-.8pc}
  \label{fig:timeline}
\end{figure}

%\subsection{User Active Time}
We also analyzed the user active time, {\em i.e.}, the time that \textbf{users spent on interacting with the system}.
Even though it is expected that \system requires more time since the user needs to talk with the crowd, 
it is still important to understand how much time it takes a user to create a rule.
In our study, participants spent an average of 2 minutes and 45 seconds (SD=1:23) to create a rule from scratch using the rule editor (``User Only'').
When using \system, participants spent an average of 3 minutes 45 seconds (SD=2:01) to converse with the crowd, and then the system took about one minute after the conversation to create a rule that the participants were willing to pick (``Crowd Only'').
If the participant decided to edit the crowd-created rules he/she just picked, it took about 2 minutes for the participants to further edit the rule (``Crowd+User'').
It took approximately 20 minutes for \system to receive the rules from all 10 workers and calculate the final rule  (``Crowd Voting'').
The complete timeline is shown in Figure~\ref{fig:timeline}.
To put these numbers in context, 
a study focusing on instant messaging within small groups showed that, on average (Least-Square Means), students respond to an instant message in 32 seconds, and people in startups respond in 105 seconds~\cite{avrahami2008waiting}.

%the average response time in instant messaging is 24 seconds~\cite{isaacs2002character}. 
%24.5\% of instant messaging chats get a response within 11-30 seconds,
%and 8.2\% of the messages have longer response times~\cite{baron2010discourse}.

%\kenneth{This is new. Proofreading needed.}
On average, the ``Crowd Voting'' setting took a user one more minute than that of the ``User Only'' setting.
That is to say, with the cost of a slightly longer user active time, \system opens up a hand-free manner of creating IF-THEN rule via conversations with the crowd. We believe this is reasonable because an advantage of a speech interface is that it can be hands-free and so users can intersperse other activities while conversing to create their rules.
According to our technical evaluation, the resulting rules from \system is as high-quality as user-created rules.
It is also noteworthy that user's cognitive load when editing a rule manually and when talking with a conversational partner are very different.
When having a conversation with \system, users are free to browse the Internet, chat with other people, or even watch a video at the same time.
In our post-study survey, which we will describe in Section~\ref{sec:user-feedback}, the participants who preferred using \system over rule editor claimed that \system is ``faster'' or ``quick'', while their user active time of using \system is actually longer.

\section{Qualitative Results}
In addition to the technical evaluation, we also collected qualitative feedback about \system from participants.
This result suggests that \system provides an easier way to compose applications for the users who have difficulty creating complex rules manually on their phones.

%The users who had a hard time using the rule editor prefer to use \system.

\subsection{Feedback from Participants}
\label{sec:user-feedback}

%\kenneth{text in this section is pretty much the same, but I adjust the order. Need some proofreading.}
We collected participants' subjective feedback immediately after they finished the lab-based study.
%More people prefer editing
We asked participants what method they preferred, {\em i.e.}, \system (``Crowd+User'') or rule editor (``User Only''), and grouped them into two groups according to their preference. The feedback we received was that 4 participants preferred \system, 7 participants preferred the rule editor, and 1 participant had no preference.
%Difficulty level is similar
We also asked participants to rate the difficulty of using \system versus using the rule editor themselves, on a Likert scale, where 1 corresponds to very easy, 2 to easy, 3 to slightly easy, 4 to neither easy nor hard, 5 to slightly hard, 6 to hard and 7 corresponds to very hard. 
As shown in Table~\ref{tab:user-analysis}, compared to the participants who preferred the rule editor, the participants who preferred \system had a much higher difficulty rating for using the rule editor.
The correlation coefficient between a user's ``difficulty rating on the rule editor'' and ``preferring \system'' (prefer=1, not prefer=0) is 0.65, which is a strong correlation.
Namely, the users who had a hard time using the rule editor prefer to use \system.
A similar relation was not found between ``user's difficulty rating on \system'' and ``preferring the rule editor'' (correlation coefficient = 0.06).
Table~\ref{tab:user-analysis} also shows that the participants who preferred \system also took longer than the other group to manually compose an IF-THEN rule on average.
This result suggests that \system provides an easier ways to create IF-THEN rules for the users who have difficulty creating complex rules manually on mobile phones.
%an alternative easier method to compose applications.
The one participant who had no preference between using the rule editor and using \system gave the following feedback:
``it depends on different situations. for example: i would like to create rules through conversations with the system while driving.''
Although we recruited users without programming experience, they were somewhat tech-savvy;
these results suggest we might see an even stronger effect if \system was used by people even less comfortable with using their smartphone.

%\kenneth{what about 8}
We also asked \textit{why} participants prefer \system.
%Strong 1: Feel faster!
Interestingly, 3 out of these 4 participants said that \system is ``faster'' or ``quick'', while they actually spent longer time to create a rule via \system when comparing to the time it took them when using the rule editor.
This could be because the difficult parts of creating rules is outsourced to the crowd when using \system, and the participants do not need to develop a rule from scratch. 
Some participants also stated that \system is more flexible since it allows the user to choose from a set of rules which is sent from multiple crowd workers.
One participant who chose to use speech input said it is ``faster'' because he/she ``doesn't like to type.''

In the post-study questionnaire, we also asked participants \textit{when} they would prefer to use \system, and when they would use the rule editor.
In their responses we found that people tend to create rules via conversation when
1) the rule would be too complex, and 2) they are busy or having a tight schedule.
6 out of 12 participants said they would choose \system when the rule they want to create has too many conditions or complex logic, e.g., 
``...I cannot figure out a proper logic to state `If' and `Then', I may relay the conversation to ask help from a server.'';
3 out of 12 participants said they would choose \system when they are busy, e.g., ``I would use it when I am busy.''

% Please add the following required packages to your document preamble:
% \usepackage{booktabs}
% \usepackage{multirow}
\begin{table*}[t]
\centering
\footnotesize
\begin{tabular}{@{}cccc@{}}
\toprule
\multicolumn{2}{c}{\multirow{2}{*}{}} & \multicolumn{2}{c}{\textbf{Participants Grouped by Preference}} \\ \cmidrule(l){3-4} 
\multicolumn{2}{c}{} & \textbf{Prefer InstructableCrowd} & \textbf{Prefer Rule Editor} \\ \midrule
\multicolumn{2}{c}{\textbf{No. of participants}} & 4 & 7 \\ \midrule
\multirow{2}{*}{\textbf{\begin{tabular}[c]{@{}c@{}}Avg. Difficulty Rating\\ \end{tabular}}} & \textbf{Crowd+User} & \textbf{3.25} (SD=1.50) & 3.57 (SD=1.62) \\ \cmidrule(l){2-4} 
 & \textbf{User Only} & 4.25 (SD=1.71) & \textbf{2.29} (SD=0.76) \\ \midrule
\multicolumn{2}{c}{\textbf{\begin{tabular}[c]{@{}c@{}}Avg. Time to Create a Rule (User Only)\\ ( mm:ss ) \end{tabular}}} & 03:15 (SD=01:20) & 02:30 (SD=0:45) \\ \bottomrule
\end{tabular}
\caption{The average difficulty ratings and rule composing time of participants that prefer \system v.s. rule editor. Difficulty rating ranged from 1 (very easy) to 7 (very hard).
The participants who preferred \system had a higher difficulty rating for using the rule editor, and also took longer to manually compose a rule.
}
\label{tab:user-analysis}
\end{table*}

\begin{comment}

%\oscar{why from 1-7? is it arbitrary? or are you using a 7 point Likert scale?}
%\kenneth{yes, 7-point likert scale}
The average difficulty level of using \system was 3.5 (SD=1.4) and that of using the rule editor was 3.0 (SD=1.4).

% Positive side of rule editor (negative feedback)
Seven participants preferred using the rule editor.  %also provided feedback.
The majority of their opinion is that using rule editor is faster.
%that using rule editor is ``Faster, and more straight forward.''
6 out of 7 participants mentioned the speed, e.g., 
``(By using the rule editor) I don't have to wait for the solutions.'', and
``conversation is long. you need to edit afterwards.''
Two participants said the rule editor is simple enough to use and ``the UI basically covered all the rules I needed.''

% user-created scenarios

One single participant had no preference between using the rule editor and using InstructableCrowd. This participant gave the following feedback:
``it depends on different situations. for example: i would like to create rules through conversations with the system while driving.''
\end{comment}

%\kenneth{Add summary maybe.}

%\kenneth{add realistic rating to scenario section}

%Finally, users give use some suggestion about interface changing, and some particiapnts mentioning that the crowd is trying to push him to say something while he was still typing. that's some thing very interesting in our case.

%Also, ASR is not as bad as we think. The participanmts with ASR soens't have kti

%11 out of 12 via text rather than conversation. Maybe don't say this...

\subsection{Information Inquiry, Confirmation and Suggestions in Conversations}

%\kenneth{some changes here too. need proof reading}

We analyzed the conversations between the participants and the crowd, and found that the responses from the crowd were often requests for more information or explicit confirmations of user's intent.
Both are known to be common dialogue acts of conversational agents~\cite{walker2001date}.
%\jpb{can you quantify this?}
%\kenneth{DATE is a complex schema. Don't feel like we have time to do this...}

%The three main types of follow-up responses from the crowd are 1) \textbf{information inquiry}, 2) \textbf{confirmation}, and 3) \textbf{clarification question}.
Most of the conversations between users and the crowd is for collecting information.
For instance, in the following conversation of S3, crowd workers ask for the information that is required in order to complete the rule they are creating:

\begin{choruschat}
    \crowd{Hi, what can I help you with?}
    \user{it was snow last night and I was late for work and missed an important meeting this morning.}
    \crowd{Would you like a weather alert?}
    \crowd{What would you like us to do?}
    \user{I missed an important meeting at 9am.}
    \crowd{What time do you usually wake up?}
    \user{7am}
    \crowd{Would you like to wake up earlier if it snows? Is 1 extra hour enough?}
    \user{sure.}
\end{choruschat}

%\textbf{376}
%requester	if i have a big dinner on my calendar and i am going to be late (if i am still far away in 30 minutes), send my wife a message saying :" i might be home late") and call the florist to prepare a small bouquet.	353626078399016	2016-04-07 19:29:01
%crowd	What time might this dinner start?	A3UV55HC87DO9C	2016-04-07 19:30:06
%requester	it depends on my calendar.	353626078399016	2016-04-07 19:30:43
%crowd	Ok I'll make note to notify your wife you may be home late in any case then	A3UV55HC87DO9C	2016-04-07 19:33:14
%crowd	Also made sure to notify the florist	A3UV55HC87DO9C	2016-04-07 19:35:55
%requester	thanks	353626078399016	2016-04-07 19:36:25
%IF(s)
In the following conversation of S6, a crowd worker was trying to figure out the time of the dinner:

%conversation 376
\begin{choruschat}
    \user{if i have a big dinner on my calendar and i am going to be late (if i am still far away in 30 minutes), send my wife a message saying :" i might be home late") and call the florist to prepare a small bouquet.}
    \crowd{What time might this dinner start?}
    \user{it depends on my calendar.}
  %\item \textbf{user:} 
%\vspace{-.4pc}  \item \textbf{crowd: }
%\vspace{-.4pc}  \item \textbf{user:} 
%\vspace{-.4pc}  \item \textbf{crowd: What would you like us to do?}
%\vspace{-.4pc}  \item \textbf{user:} I missed an important meeting at 9am.
%\vspace{-.4pc}  \item \textbf{crowd: What time do you usually wake up?}
%\vspace{-.4pc}  \item \textbf{user:} 7am
%\vspace{-.4pc}  \item \textbf{crowd: Would you like to wake up earlier if it snows? Is 1 extra hour enough?}
%\vspace{-.4pc}  \item \textbf{user:} sure.
\end{choruschat}

%\begin{comment}

%\kenneth{this can be removed if not necessary}
In the following conversation of a different user for the same scenario S6, a different crowd worker asked similar follow-up questions:

\begin{choruschat}
    \user{I don't want to be late for home too often, otherwise my wife would get angry at me}
    \crowd{So how may I help you}
    \crowd{when do you want to get an alert?}
    \user{can you send Amy a message saying I might be home late}
    \user{yes}
    \crowd{what time do you want this to be sent?}
    \user{if I'm going to be late}
    \crowd{what time is late?}
    \user{for our scheduled dinner on my calendar}
\end{choruschat}

Crowd workers sometimes confirmed with the users information which was conveyed previously.
For example, in the following conversation of S5, a worker asked a confirmation question about the time.

%\textbf{495}
%crowd	hello?	A19BMRBO5UAD6H	2016-04-08 13:28:50
%requester	I leave work after 5pm and take Bus 53 home at the Washington street	353626078399016	2016-04-08 13:29:07
%requester	I don't wanna wait for the bus for too long unless the bus is coming soon	353626078399016	2016-04-08 13:29:32
%requester	let me know the schedule for the bus 53 ten minutes ahead	353626078399016	2016-04-08 13:30:26
%crowd	is after 5pm	A2CCSFYX3X4ECB	2016-04-08 13:30:42
%requester	yes	353626078399016	2016-04-08 13:30:48
%crowd	ok	A2CCSFYX3X4ECB	2016-04-08 13:30:56
%crowd	do you need email or message or call	A2CCSFYX3X4ECB	2016-04-08 13:31:15
%crowd	yes surely willl do that	A8WLCSZLNZ8BL	2016-04-08 13:31:35

\begin{choruschat}
    \crowd{hello?}
    \user{I leave work after 5pm and take Bus 53 home at the Washington street}
    \user{I don't wanna wait for the bus for too long unless the bus is coming soon}
    \crowd{is after 5pm}
    \user{yes}
    %\crowd{do you need email or message or call}
    %\crowd{}
  %\item \textbf{crowd: } 
%\vspace{-.4pc}  \item \textbf{user:} 
%\vspace{-.4pc}  \item \textbf{user:} 
%\vspace{-.4pc}  \item \textbf{crowd:}
%\vspace{-.4pc}  \item \textbf{user:} 
%\vspace{-.4pc}  \item \textbf{crowd: ok}
%\vspace{-.4pc}  \item \textbf{crowd: }
%\vspace{-.4pc}  \item \textbf{crowd: Would you like to wake up earlier if it snows? Is 1 extra hour enough?}
%\vspace{-.4pc}  \item \textbf{user:} sure.
\end{choruschat}

Furthermore, open conversation can lead to solutions that the user did not think of.
For example, in the following conversation of S2, the crowd worker suggested to send a message back or to use an alarm/notification, instead of setting a phone call.
The alternatives that the crowd came up with demonstrates their potential to be creative and think of solutions that the user might not have.
%instead of a phone call
%, though the user did not accept this suggestion.

%\textbf{354}

%crowd	Hello, how can I help you?	A2UAP3YWCX2NHN	2016-04-07 17:42:24
%requester	please call me if the text from my mom containing "grandpa" or"grandfather".	353626078399016	2016-04-07 17:44:44
%crowd	Do you want to send them a message asking to call you, or do you want to receive an alarm or notification?	A2UAP3YWCX2NHN	2016-04-07 17:46:31
%requester	maybe just call me. thanks!	353626078399016	2016-04-07 17:47:22

\begin{choruschat}
    \crowd{Hello, how can I help you??}
    \user{please call me if the text from my mom containing ``grandpa'' or ``grandfather''.}
    \crowd{Do you want to send them a message asking to call you, or do you want to receive an alarm or notification?}
    \user{maybe just call me. thanks!}
\end{choruschat}

\subsection{Alternative Solutions for the Same Scenario}
%\kenneth{not sure we need this}
%\amos{Maybe move part of this above, when we say that we accepted rules as correct even if they were different that what we have originally thought about}
%\kenneth{but I think this has too much details. maybe keep it here is ok.}

%\kenneth{proofreading needed.}

We observed that participants and workers could come up with different rules in response to a same scenario, for four main reasons:
First, people have their own preferred ways to be notified under different circumstances, and thus sometimes chose different effectors than we intended in their rules.
For instance, more than one participant tried to add extra effectors, such as an alarm in the ``Message'' scenario (S2.) because they believed missing a message about the hospitalized grandfather can be quite serious.
Second, similarly, users also have their own preferences for sensors.
For example, in the ``Snow \& Meeting'' scenario (S3,) one participant selected ``News'' in addition to the gold-standard sensors and argued that s/he would only wake up for heavy snow, which is likely to be mentioned in the news.
Third, some alternative rules created by crowd workers may be caused by the ambiguities in user's instruction.
%instructions from the user.
%Amos says: Kenneth, is this correct? I edited this but wasn't sure what you meant.
%in our scenarios, the THEN (action) part of a rule is often more flexible than the IF (condition) and thus allows more alternative solutions.
For instance, in the following conversation of S4, the word ``reply'' does not necessarily imply ``sending a message'' (although it might be the most common solution).
Therefore, ``sending an email'' is also acceptable. 

%\textbf{336}
%Role	Chat	Worker ID	Time
%requester	hi	353626078399016	2016-04-07 17:12:22
%requester	I know car accidents might happen if i talk on the phone while driving. so I would like %to reply "sorry I am driving" to anyone calling me when I'm driving.	353626078399016	2016-04-07 %crowd ok i will do so now	A8WLCSZLNZ8BL	2016-04-07 17:15:46

%conversation 316
\begin{choruschat}
    \user{hi}
    \user{I know car accidents might happen if i talk on the phone while driving. so I would like to reply ``sorry I am driving'' to anyone calling me when I'm driving.}
    \crowd{ok i will do so now}
\end{choruschat}

Finally, %Multiple effective but different rules were sometimes created for the same scenario. 
sometimes two different rules can behave similarly or even identically in the real world.
For example, in the ``Bus'' scenario (S5), the notification can either be fired when ``the Bus 53 will arrive at Washington St. in 5 miniutes'' or when ``Bus 53 is arriving at Hamilton St. stop now,'' since the Bus 53 usually takes 5 minutes to drive from Hamilton St. stop to Washington St.
Both rules occurred in our study.

%------------------

%Some other users also want to add alarm of phone call to the Grandpa senario because he really being afraid of missing important information form mom.

\section{Discussion}
In this paper we introduced \system, a system that allows users to create IF-THEN rules for smartphones via conversation with the crowd. This work provides a potential route toward more interesting conversations with intelligent agents than is currently possible. In this section we discuss some of the issues and reflections that came from the development and study of \system.

\subsection{Assessing Performance and Goal Achievement}

The study showed that the performance of the crowd system is nearly the same as that of a typical GUI in terms of the \textit{quality} of the generated rules. This might lead some to question whether users would want to use \system if it is not \textit{better} than other options at creating accurate IF-THEN rules.
The motivation of \system is to challenge the traditional methods of manually composing an IF-THEN rule within the context of performing complex tasks via alternative interfaces. The key question we wanted to answer is: ``Can the system perform as well as users themselves, while employing a new method of doing it?''. Outsourcing complex tasks to the crowd is not always about whether or not the system can do it ``better.'' Often it is about opening up an opportunity to achieve the same goal using a different technology or method, in this case via natural language interface. In this respect, ``better'' really depends on how one is assessing achievement of the goal.
In prior projects within this theme, crowd-powered systems have not always performed \textit{better} than users. In WearWrite~\cite{WearWrite}, Chorus~\cite{Chorus}, and Knowledge Accelerator~\cite{KA,Alloy}, the proposed solutions did not necessarily produce results that were faster or of higher quality than traditional methods. The value of these projects was opening up new possibilities of completing tasks in ways that were not possible before, especially with respect to flexibility. Creating a blog post by talking to a smartwatch with WearWrite will not necessarily result in higher article quality than typing it on a laptop, but the system lets users create content on the fly nearly anywhere. Searching via Chorus crowd workers might not provide better results than just using a search engine, but it is much more convenient. Similarly, Knowledge-Accelerator's use of crowd workers allows a user to  ask an open-ended question and get a sophisticated answer in few hours, and open-ended questions are something that computers do not deal with very well.

\subsection{Challenges in Producing High-Quality Rules}

Creating a multi-part IF-THEN rule is difficult because computer-executable rules (like all programs) have little tolerance for mistakes. If we break down an IF-THEN rule to a composition of sensors and effectors with attribute values, experiments have shown that humans are reasonably good at composing sensors/effectors and filling their attributes, respectively. However, when we add up all the work, any mistakes will make the resulting IF-THEN rule ineffective. A natural response to issue would be to enforce stricter validation for human input in rule creation. However, strict input validation on the interface would increase the time it takes to create a rule for both users and the crowd and frustrate users more easily. It would also  increase the engineering effort required to add a new sensor or effector to the system, which often come with arbitrary constraints. IFTTT, as a successful rule-creation product, avoids multiple sensors and effectors, and uses a user-friendly workflow to balance possible user frustration. Our project suggests using conversation and iterative editing to permit robust rule creation.

\subsection{Rule Validation}

One of the most common issues faced by the Decision-Rule Engine is the \textit{rule conflict resolution} issue: deciding which rule should be triggered when there are multiple with the same set of conditions (IFs) but a different set of actions (THENs).
If a user receives multiple rules during the same conversational session, it is reasonable to assume that they are redundant and to allow the user to pick only a single rule from this set. However, if the user creates many rules in many different sessions, he or she may forget about a created rule and attempt to create the same rule again. Furthermore, the user may at first create a very specific rule (e.g., IF I have a meeting at 9 am, THEN notify me the night before) and later try to generalize it (e.g., IF I have a meeting at 10 am or earlier, THEN notify me the night before).
If the Decision-Rule Engine were to follow these rules regardless of conflicts, the same action might be executed more than once, which is not likely the user's intent. Currently, the Monitoring/Tracking module may detect these kinds of conflicts and automatically subsume the less-used rules, but further research is required into identifying these cases and alerting the user in advance.
One approach could be keeping these conflicting rules and defining some heuristics that would determine when a rule should subsume or inhibit others, or when they should be executed sequentially, etc. Another approach could be defining a mechanism that removes those rules that are redundant or conflicting and less relevant than others (with the user's approval).

\subsection{Timing of Executing Triggers and Actions}
Different sensors and effectors may require very different frequencies.
For example, while a weather-related sensor trigger such as ``IF it is snowing early in the morning'' can be checked once every 24 hours, a ``phone body'' sensor trigger connected to the phone's accelerometer (e.g., ``IF the phone is dropping towards the floor'') might need to be checked every 100 milliseconds.
Other sensor conditions, such as calendar events ({\em e.g.} ``IF I have a meeting tomorrow before 10am'') may be validated immediately after the rule is created, and then checked again every hour (in case new meetings have been added).
Similarly, effectors also have different execution timing requirements. Some actions can be executed immediately after the conditions are met, while others must be scheduled for later execution. For instance, the action ``THEN show me a notification right now'' is executed right after the conditions are fulfilled, whereas the action ``THEN send me a reminder tonight at 10 pm'' would be scheduled for execution at the appropriate time.
Currently, the Rule Validator in \system's middleware uses different timing validation mechanisms for different sensors and effectors. To scale up to a larger number of sensors and effectors, a more systematic manner for categorizing the frequency ranges of sensors and effectors is likely required.

\subsection{User Privacy}

One participant in our study expressed a concern about user privacy.
In the current prototype, a limited view of a user's personal information ({\em e.g.}, a contact list created for the purpose of the study) was exposed to crowd workers. In the future, we may use aliases that are either automatically assigned or created by the user to prevent true names or other information from being disseminated to crowd workers. For instance, instead of an actual address, the user could provide an alias such as ``Home'' or ``Office'' when talking to the crowd. Aliases can also be used to protect information about people or time ({\em e.g.}, using ``Wife'' instead of ``Amy,'' or ``Birthday'' instead of the actual date.) However, the use of aliases cannot completely prevent the user from providing personal information in conversation.
While several privacy-preserving human computation workflows have been proposed for annotating videos~\cite{lasecki2015exploring} and accessing users' personal information~\cite{wearmail2017}, privacy is still a well-known issue in the field of crowdsourcing, since the data is processed by human workers.
A future direction is to further explore privacy issues that may arise with conversational interfaces.

\subsection{Limitations}

One natural limitation of the architecture of \system is that all the sensors and effectors must be \textit{comprehensible} to the majority of crowd workers. For example, despite being one of the most common built-in sensors in smartphones, the accelerometer sensor's raw output is very difficult to use directly by non-experts to interpret certain movements of the phone ({\em e.g.}, falling or being in motion while driving or walking). Future systems may find value in explicitly recruiting to their crowds people with programming expertise who can provide abstractions of raw sensor values that could be shared and reused by others.
Using current sensors to express high-level semantics ({\em e.g.}, determining when the user is sleeping) requires specialized knowledge that most crowd workers likely do not have. 
IF-THEN rules have low tolerance for mistakes, and quality control is still an essential challenge in crowdsourcing.
It may be useful to explore ways for the rules that are created to form a part of a probabilistic suggestion system, {\em i.e.}, instead of automatically conducting an action that may or may not be correct, ask the user whether or not to do it.

\section{Future Work}
\system suggests a number of opportunities for future work. %In this section 
With the help of crowd workers, \system is able to convert a natural language conversation to an IF-THEN rule.
Human workers are known to be able to perform various tasks that automated systems still can not do, however, often with the cost of longer latency and higher operating budget.
One natural follow-up step is to explore the potential of automating the process of \system.
While the automated approach did not perform as well as humans in prior work, a better performance can be expected when the system is able to collect larger amount of training data.
Furthermore, the attribute filling task in creating IF-THEN rules is similar to the ``slot filling'' task in dialogue systems, in which we can take advantage of existent approaches such as Conditional Random Fields (CRF)~\cite{AtisSpain} or Recurrent Neural Networks (RNN)~\cite{mesnil2015using}.
Creating multi-part IF-THEN rules is a challenging task, for both human and machines.
We imagine a future that automated components can work with human workers to make such systems more robust and scalable.

\begin{figure}[t]
  \centering
  %\small
  \includegraphics[width=0.95\columnwidth]{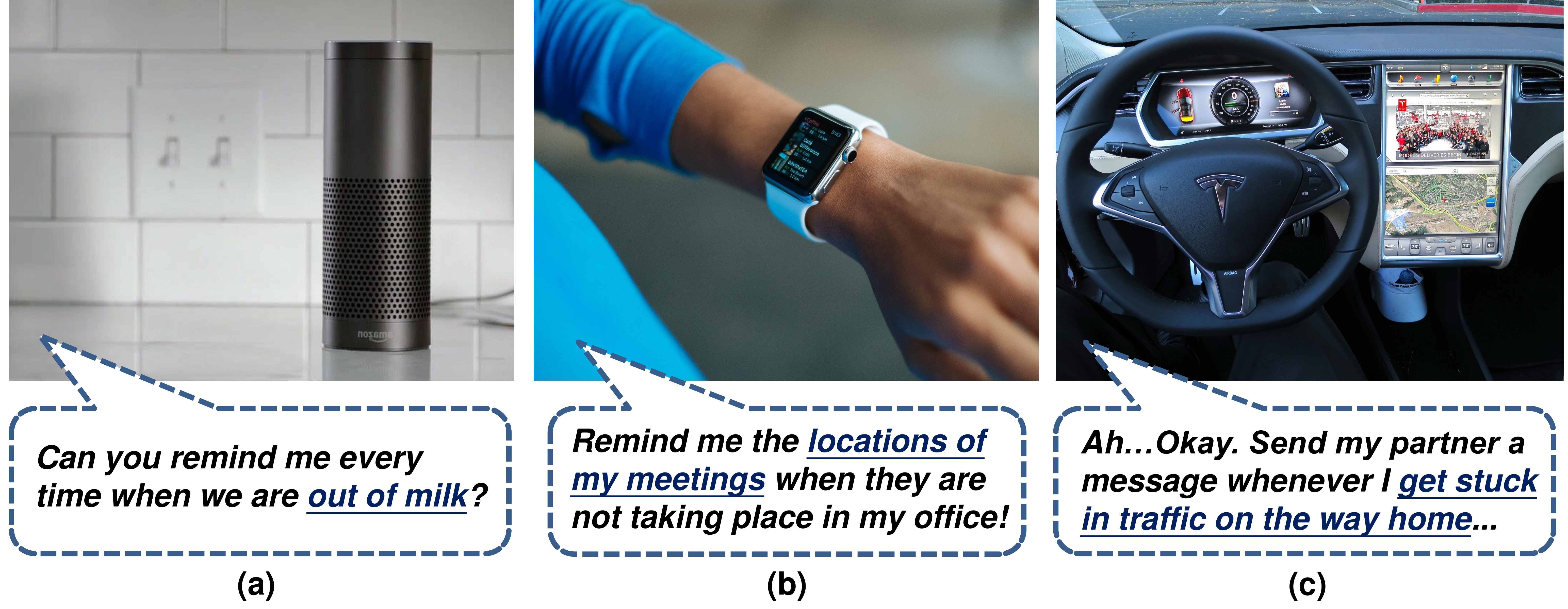}
  \caption{Scenarios of future conversational assistants that allow end-users to verbally create IF-THEN rules to control smart devices. When end-users experience a problem such as \emph{(a)} being out of milk, \emph{(b)} forgetting the meeting room, or \emph{(c)} getting stuck in traffic when driving home, they can verbally instruct their assistants at the scene to set up IF-THEN rules to prevent the problems from happening again. The framework of \system can not only implement on the mobile phone, but also smart watch and voice-enabled devices such as Amazon's Echo.}
  \label{fig:vision}
\end{figure}

Furthermore, \system introduced a new interaction paradigm of conversational agents, which can not only be implemented in smartphones, but also be applied to smart homes, smart watches, voice-enabled devices such as Amazon's Echo, or smart cars in hand-free scenarios.
End-users can freely record the problems the are experiencing and create an IF-THEN rule to solve it via any devices that are available at the spot.
Figure~\ref{fig:vision} illustrates potential user scenarios of future \system on different devices.
In a smart home setting, when the user open the smart refrigerator and find that they are out of milk, he/she can tell their Echo in the kitchen to create a rule that reminds them they do not have much milk left in the refrigerator (Figure~\ref{fig:vision} (a));
when a professor realizes that the next meeting will not be held in his/her own office but can not remember the room, he/she can set up a rule be talking to the smartwatch to set up a push notification about the room if the incoming meeting is in a different room;
and when users get stuck in traffic when driving home, they can talk to the smart car panel and set up an automatic message whenever they will be late home (Figure~\ref{fig:vision} (c)).
Voice interface opens up many possibilities of end-users to keep track of their behavior and improve life quality in the moment, and we believe that enabling users to create IF-THEN rules by talking to their smartphones is a promising start.

\section{Conclusion}
In this paper we introduced \system, a system that allows end users to create complex IF-THEN rules via voice in collaboration with the crowd.
%converse with the crowd via smart phone, and have the crowd create an IF-THEN rule to help with the user's problem.
These rules connect to the sensors and the effectors on the user's phone where the sensors serve as triggers and the effectors as actions. %to help them get things done.
We have built support for crowd workers to have a conversation with the users and allow them to suggest rules for the users. 
%define these rules for them.
A user study shows that non-programmers can effectively create rules via conversation, and suggests that collaboration between the user and the crowd while creating IF-THEN rules could be a fruitful area for future research.
%We suggest a number of different avenues for further refinement going forward.
%Our continued work will focus on enabling robust creation of IF-THEN rules via voice.
%We will iterate on our approach for allowing the crowd to create IF-THEN rules via conversation using the sensors and effectors available through the middleware.
%We will evaluate this first in controlled studies and then in the deployed agent.
As we collect examples of IF-THEN rules, we will look for ways to use them to automate the creation of common IF-THEN patterns.
%Users will want feedback when collaborating with the systems to have confidence that what they intended to create was actually created.%Amos Says: I think that this was an old sentence, we do give them feedback in the new system
%We will work with users to develop appropriate ways for them to validate and edit the rules that are created.
%This work will contribute to research in human computation and end user programming.
%We are also looking forward to cooperating with more applications.
More broadly, \system represents a new approach in which end users work with remote crowd workers to bring about powerful functionality despite the constraints of mobile and wearable devices.

%\oscar{We have created a generic JSON representation for mapping sensors onto rule conditions and effectors onto rule actions. We implemented a novel Decision-Rule Engine whose purpose is many-fold: 1) to parse json structures coming from the crowd-sourcing web application into executable decision rules that run in the smartphone; 2) to validate whether a rule should be triggered because all their conditions are fulfilled; 3) to execute rules at different timing periods; 4) to maintain a knowledge base; 5) to monitor the execution of rules in order to find rule conflicts; and 6) to interface with sensors, effectors and services.}

%We have created a generic JSON representation for the sensors and the effectors in a smartphone, and the ability to define rules. We created a middleware that allows access to these sensors and effectors and applies these rules.

\section{Acknowledgements}
This work was funded by the National Science Foundation under Award \#IIS-1816012, and as part of the Yahoo! InMind project. We thank the workers on Amazon Mechanical Turk for making this work possible.

\bibliography{hcj-sample}

\end{document}